\documentclass[journal,twocolumn,10pt,twoside]{IEEEtran}
\normalsize

\ifCLASSINFOpdf
\else
\fi

\usepackage{amsmath}
\usepackage{amssymb}
\usepackage{cite}
\usepackage{graphicx}
\usepackage{algorithm}
\usepackage{algpseudocode}
\usepackage{subfigure}
\usepackage{multirow} 
\usepackage{longtable}
\usepackage{threeparttable}
\usepackage{caption3}
\usepackage{array}
\usepackage{cases}
\usepackage{color}
\usepackage[T1]{fontenc}
\usepackage[utf8]{inputenc}
\usepackage{authblk}
\usepackage{subfigmat}

\newtheorem{remark}{Remark}

\newtheorem{observation}{Observation}

\newcommand\mycom[2]{\genfrac{}{}{0pt}{}{#1}{#2}}

\begin{document}
\title{Wireless Information and Power Transfer: Nonlinearity, Waveform Design and Rate-Energy Tradeoff}
\author{Bruno Clerckx 
\thanks{Bruno Clerckx is with the EEE department at Imperial College London, London SW7 2AZ, UK (email: b.clerckx@imperial.ac.uk). This work has been partially supported by the EPSRC of UK, under grant EP/P003885/1. The material in this paper was presented in part at the ITG WSA 2016 \cite{Clerckx:2016}.}}

\maketitle

\begin{abstract}
The design of Wireless Information and Power Transfer (WIPT) has so far relied on an oversimplified and inaccurate linear model of the energy harvester. In this paper, we depart from this linear model and design WIPT considering the rectifier nonlinearity. We develop a tractable model of the rectifier nonlinearity that is flexible enough to cope with general multi-carrier modulated input waveforms. Leveraging that model, we motivate and introduce a novel WIPT architecture relying on the superposition of multi-carrier unmodulated and modulated waveforms at the transmitter. The superposed WIPT waveforms are optimized as a function of the channel state information so as to characterize the rate-energy region of the whole system. Analysis and numerical results illustrate the performance of the derived waveforms and WIPT architecture and highlight that nonlinearity radically changes the design of WIPT. We make key and refreshing observations. \textit{First}, analysis (confirmed by circuit simulations) shows that modulated and unmodulated waveforms are not equally suitable for wireless power delivery, namely modulation being beneficial in single-carrier transmissions but detrimental in multi-carrier transmissions. \textit{Second}, a multi-carrier unmodulated waveform (superposed to a multi-carrier modulated waveform) is useful to enlarge the rate-energy region of WIPT. \textit{Third}, a combination of power splitting and time sharing is in general the best strategy. \textit{Fourth}, a non-zero mean Gaussian input distribution outperforms the conventional capacity-achieving zero-mean Gaussian input distribution in multi-carrier transmissions. \textit{Fifth}, the rectifier nonlinearity is beneficial to system performance and is essential to efficient WIPT design.
\end{abstract}

\begin{IEEEkeywords} Nonlinearity, optimization, waveform, wireless power, wireless information and power transfer
\end{IEEEkeywords}

\IEEEpeerreviewmaketitle

\vspace{-0.3cm}
\section{Introduction}


\par Wireless Information and Power Transfer/Transmission (WIPT) is an emerging research area that makes use of radiowaves for the joint purpose of wireless communications or Wireless Information Transfer (WIT) and Wireless Power Transfer (WPT). WIPT has recently attracted significant attention in academia. It was first considered in \cite{Varshney:2008}, where the rate-energy tradeoff was characterized for some discrete channels, and a Gaussian channel with an amplitude constraint on the input. WIPT was then studied in a frequency-selective AWGN channel under an average power constraint \cite{Grover:2010}. Since then, WIPT has attracted significant interests in the communication literature with among others MIMO broadcasting \cite{Zhang:2013,Son:2014,Xu:2014}, architecture \cite{Zhou:2013}, interference channel \cite{Park:2013,Park:2014,Park:2015}, broadband system \cite{Huang:2013,Zhou:2014,Ng:2013}, relaying \cite{Nasir:2013,Huang:2015,Huang:2016r}, wireless powered communication \cite{Ju:2014,Lee:2016}. Overviews of potential applications and promising future research avenues can be found in \cite{Lu:2015,KHuang:2016}. 

\par Wireless Power Transfer (WPT) is a fundamental building block of WIPT and the design of an efficient WIPT architecture fundamentally relies on the ability to design efficient WPT. The major challenge with WPT, and therefore WIPT, is to find ways to increase the end-to-end power transfer efficiency, or equivalently the DC power level at the output of the rectenna for a given transmit power. To that end, the traditional line of research (and the vast majority of the research efforts) in the RF literature has been devoted to the design of efficient rectennas \cite{OptBehaviour,Costanzo:2016} but a new line of research on communications and signal design for WPT has emerged recently in the communication literature \cite{Zeng:2017}.
 
\par A rectenna is made of a nonlinear device followed by a low-pass filter to extract a DC power out of an RF input signal. The amount of DC power collected is a function of the input power level and the RF-to-DC conversion efficiency. Interestingly, the RF-to-DC conversion efficiency is not only a function of the rectenna design but also of its input waveform (power and shape) \cite{Trotter:2009,Boaventura:2011,Collado:2014,Valenta:2015,Clerckx:2015,Clerckx:2016b,Boshkovska:2015}. This has for consequence that the conversion efficiency is not a constant but a \textit{nonlinear} function of the input waveform (power and shape).
 
\par This observation has triggered recent interests on systematic wireless power waveform design \cite{Clerckx:2016b}. The objective is to understand how to make the best use of a given RF spectrum in order to deliver a maximum amount of DC power at the output of a rectenna. This problem can be formulated as a link optimization where transmit waveforms (across space and frequency) are adaptively designed as a function of the channel state information (CSI) so as to maximize the DC power at the output of the rectifier. In \cite{Clerckx:2016b}, the waveform design problem for WPT has been tackled by introducing a simple and tractable analytical model of the \textit{diode nonlinearity} through the second and higher order terms in the Taylor expansion of the diode characteristics. Comparisons were also made with a linear model of the rectifier, that only accounts for the second order term, which has for consequence that the harvested DC power is modeled as a conversion efficiency constant (i.e.\ that does not reflect the dependence w.r.t. the input waveform) multiplied by the average power of the input signal.
Assuming perfect Channel State Information at the Transmitter (CSIT) can be attained, relying on both the linear and nonlinear models, an optimization problem was formulated to adaptively change on each transmit antenna a multisine waveform as a function of the CSI so as to maximize the output DC current at the energy harvester. Important conclusions of \cite{Clerckx:2016b} are that 1) multisine waveforms designed accounting for nonlinearity are spectrally more efficient than those designed based on a linear model of the rectifier, 2) the derived waveforms optimally exploit the combined effect of a beamforming gain, the rectifier nonlinearity and the channel frequency diversity gain, 3) the linear model does not characterize correctly the rectenna behavior and leads to inefficient multisine waveform design, 4) rectifier nonlinearity is key to design efficient wireless powered systems. Following \cite{Clerckx:2016b}, various works have further investigated WPT signal and system design accounting for the diode nonlinearity, including among others waveform design complexity reduction \cite{Huang:2016,Huang:2017,Clerckx:2017,Moghadam:2017}, large-scale system design with many sinewaves and transmit antennas \cite{Huang:2016,Huang:2017}, multi-user setup \cite{Huang:2016,Huang:2017}, imperfect/limited feedback setup \cite{Huang:2017b}, information transmission \cite{Kim:2016} and prototyping and experimentation \cite{Kim:2017}. Another type of nonlinearity leading to an output DC power saturation due to the rectifier operating in the diode breakdown region, and its impact on system design, has also appeared in the literature \cite{Boshkovska:2015,Xu:2017}.

\par Interestingly, the WIPT literature has so far entirely relied on the linear model of the rectifier, e.g. see \cite{Varshney:2008,Grover:2010,Zhang:2013,Son:2014,Xu:2014,Zhou:2013,Park:2013,Park:2014,Park:2015,Huang:2013,Zhou:2014,Ng:2013,Nasir:2013,Huang:2015,Huang:2016r,Ju:2014,Lee:2016,Lu:2015,KHuang:2016}. Given the inaccuracy and inefficiency of this model and the potential of a systematic design of wireless power waveform as in \cite{Clerckx:2016b}, it is expected that accounting for the diode nonlinearity significantly changes the design of WIPT and is key to efficient WIPT design, as confirmed by initial results in \cite{Clerckx:2016}.

 
\par In this paper, we depart from this linear model and revisit the design of WIPT in light of the rectifier nonlinearity. We address the important problem of waveform and transceiver design for WIPT and characterize the rate-energy tradeoff, accounting for the rectifier nonlinearity. In contrast to the existing WIPT signal design literature, our methodology in this paper is based on a bottom-up approach where WIPT signal design relies on a sound science-driven design of the underlying WPT signals initiated in \cite{Clerckx:2016b}. 

\par \textit{First}, we extend the analytical model of the rectenna nonlinearity introduced in \cite{Clerckx:2016b}, originally designed for multi-carrier unmodulated (deterministic multisine) waveform, to multi-carrier modulated signals. We investigate how a multi-carrier modulated waveform (e.g. OFDM) and a multi-carrier unmodulated (deterministic multisine) waveform compare with each other in terms of harvested energy. Comparison is also made with the linear model commonly used in the WIPT literature. Scaling laws of the harvested energy with single-carrier and multi-carrier modulated and unmodulated waveforms are analytically derived as a function of the number of carriers and the propagation conditions. Those results extend the scaling laws of \cite{Clerckx:2016b}, originally derived for unmodulated waveforms, to modulated waveforms. We show that by relying on the classical linear model, an unmodulated waveform and a modulated waveform are equally suitable for WPT. This explains why the entire WIPT literature has used modulated signals. On the other hand, the nonlinear model clearly highlights that they are not equally suitable for wireless power delivery, with modulation being beneficial in single-carrier transmission but detrimental in multi-carrier transmissions. The behavior is furthermore validated through circuit simulations. This is the first paper where the performance of unmodulated and modulated waveforms are derived based on an tractable analytical model of the rectifier nonlinearity and the observations made from the analysis are validated through circuit simulations.   
 
\par \textit{Second}, we introduce a novel WIPT transceiver architecture relying on the superposition of multi-carrier unmodulated and modulated waveforms at the transmitter and a power-splitter receiver equipped with an energy harvester and an information decoder. The WIPT superposed waveform and the power splitter are jointly optimized so as to maximize and characterize the rate-energy region of the whole system. The design is adaptive to the channel state information and results from a posynomial maximization problem that originates from the nonlinearity of the energy harvester. This is the first paper that studies WIPT and the characterization of the rate-energy tradeoff considering the diode nonlinearity.

\par \textit{Third}, we provide numerical results to illustrate the performance of the derived waveforms and WIPT architecture. Key observations are made. \textit{First}, a multi-carrier unmodulated waveform (superposed to a multi-carrier modulated waveform) is useful to enlarge the rate-energy region of WIPT if the number of subbands is sufficiently large (typically larger than 4). \textit{Second}, a combination of power splitting and time sharing is in general the best strategy. \textit{Third}, a non-zero mean Gaussian input distribution outperforms the conventional capacity-achieving zero-mean Gaussian input distribution in multi-carrier transmissions. \textit{Fourth}, the rectifier nonlinearity is beneficial to system performance and is essential to efficient WIPT design. This is the first paper to make those observations because they are direct consequences of the nonlinearity.  

\par \textit{Organization:} Section \ref{SWIPT_section} introduces and models the WIPT architecture. Section \ref{section_SWIPT_waveform} optimizes WIPT waveforms and characterizes the rate-energy region. Section \ref{scaling_laws} derives the scaling laws of modulated and unmodulated waveforms. Section \ref{simulations} evaluates the performance and section \ref{conclusions} concludes the work.
\par \textit{Notations:} Bold lower case and upper case letters stand for vectors and matrices respectively whereas a symbol not in bold font represents a scalar. $\left\|.\right\|_F^2$ refers to the Frobenius norm a matrix. $\mathcal{A}\left\{.\right\}$ refers to the DC component of a signal. $\mathcal{E}_X\left\{.\right\}$ refers to the expectation operator taken over the distribution of the random variable $X$ ($X$ may be omitted for readability if the context is clear). $.^*$ refers to the conjugate of a scalar. 
$\left(.\right)^T$ and $\left(.\right)^H$ represent the transpose and conjugate transpose of a matrix or vector respectively. The distribution of a
circularly symmetric complex Gaussian (CSCG) random vector with mean $\mathbf{\mu}$ and covariance matrix $\mathbf{\Sigma}$ is denoted by $\mathcal{CN}(\mu,\Sigma)$ and $\sim$ stands for ``distributed as''. 

\vspace{-0.1cm}
\section{A Novel WIPT Transceiver Architecture}\label{SWIPT_section}
In this section, we introduce a novel WIPT transceiver architecture and detail the functioning of the various building blocks.
The motivation behind the use of such an architecture will appear clearer as we progress through the paper.
 
\subsection{Transmitter and Receiver}
\par We consider a single-user point-to-point MISO WIPT system in a general multipath environment. The transmitter is equipped with $M$ antennas that transmit information and power simultaneously to a receiver equipped with a single receive antenna. We consider the general setup of a multi-carrier/band transmission (with single-carrier being a special case) consisting of $N$ orthogonal subbands where the $n^{\textnormal{th}}$ subband has carrier frequency $f_n$ and equal bandwidth $B_s$, $n=0,...,N-1$. The carrier frequencies are evenly spaced such that $f_n=f_0+n \Delta_f$ with $\Delta_f$ the inter-carrier frequency spacing (with $B_s\leq \Delta_f$).

\par
\begin{figure}
\begin{minipage}[c]{\columnwidth}
\centering
\includegraphics[width=0.45\columnwidth]{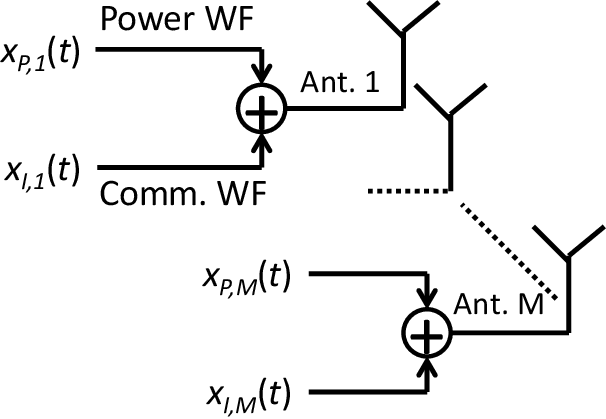}\\
\small{(a) Transmitter}
\end{minipage}\vspace{0.2cm}
\begin{minipage}[c]{\columnwidth}
\centering
\includegraphics[width=0.85\columnwidth]{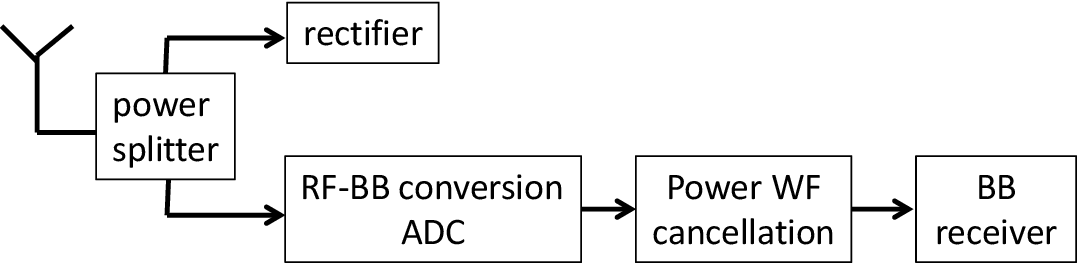}\\
\small{(b) Receiver with waveform cancellation}
\end{minipage}
\caption{Transceiver (Tx and Rx) architecture for WIPT with superposed communication and power waveform (WF).}
\label{WIPT_transceiver}
\vspace{-0.2cm}
\end{figure}

 Uniquely, the WIPT signal transmitted on antenna $m$, $x_m(t)$, consists in the superposition of one multi-carrier unmodulated (deterministic multisine) power waveform $x_{P,m}(t)$ at frequencies $f_n$, $n=0,...,N-1$ for WPT and one multi-carrier modulated communication waveform $x_{I,m}(t)$ at the same frequencies for WIT\footnote{$x_{I,m}(t)$ can be implemented using e.g. OFDM.}, as per Fig \ref{WIPT_transceiver}(a). The modulated waveform carries $N$ independent information symbols $\tilde{x}_n(t)$ on subband $n=0,...,N-1$. Hence, the transmit WIPT signal at time $t$ on antenna $m=1,...,M$ writes as
\begin{align}
x_{m}(t)&=x_{P,m}(t)+x_{I,m}(t),\nonumber\\
&=\sum_{n=0}^{N-1}s_{P,n,m} \cos(2\pi f_n t+\phi_{P,n,m})\nonumber\\
&\hspace{1.5cm}+\tilde{s}_{I,n,m}(t) \cos(2\pi f_n t+\tilde{\phi}_{I,n,m}(t)),\nonumber\\
&=\Re\left\{\sum_{n=0}^{N-1} \left(w_{P,n,m}+x_{n,m}(t)\right) e^{j 2\pi f_n t}\right\},\nonumber\\
&=\Re\left\{\sum_{n=0}^{N-1} \left(w_{P,n,m}+w_{I,n,m}\tilde{x}_n(t)\right) e^{j 2\pi f_n t}\right\}\label{SWIPT_WF}
\end{align}
where we denote the complex-valued baseband signal transmitted by antenna $m$ at subband $n$ for the unmodulated (deterministic multisine) waveform as $w_{P,n,m}=s_{P,n,m}e^{j \phi_{P,n,m}}$ and for the modulated waveform as $x_{n,m}(t)=w_{I,n,m}\tilde{x}_n(t)=\tilde{s}_{I,n,m}(t)e^{j \tilde{\phi}_{I,n,m}(t)}$. $w_{P,n,m}$ is constant across time (for a given channel state) and $x_{P,m}(t)$ is therefore the weighted summation of $N$ sinewaves inter-separated by $\Delta_f$ Hz, and hence occupies zero bandwidth. On the other hand, $x_{n,m}(t)$ has a signal bandwidth no greater than $B_s$ with symbols $\tilde{x}_n(t)$ assumed i.i.d. CSCG\footnote{following the capacity achieving input distribution in a Gaussian channel with average power constraint.} random variable with zero-mean and unit variance (power), denoted as $\tilde{x}_n\sim \mathcal{CN}(0,1)$. Denoting the input symbol $\tilde{x}_{n}\!=\!\left|\tilde{x}_{n}\right|e^{j\phi_{\tilde{x}_n}}$, we further express the magnitude and phase of $x_{n,m}$ as follows $\tilde{s}_{I,n,m}\!=\!s_{I,n,m}\left|\tilde{x}_{n}\right|$ with $s_{I,n,m}\!=\!\left|w_{I,n,m}\right|$ and $\tilde{\phi}_{I,n,m}\!=\!\phi_{I,n,m}\!+\!\phi_{\tilde{x}_n}$. Hence $\mathcal{E}\big\{\left|x_{n,m}\right|^2\big\}=s_{I,n,m}^2$ and $x_{n,m}\sim \mathcal{CN}(0,s_{I,n,m}^2)$.

\par The transmit WIPT signal propagates through a multipath channel, characterized by $L$ paths. Let $\tau_l$ and $\alpha_l$ be the delay and amplitude gain of the $l^{\textnormal{th}}$ path, respectively. Further, denote by $\zeta_{n,m,l}$ the phase shift of the $l^{\textnormal{th}}$ path between transmit antenna $m$ and the receive antenna at subband $n$. Denoting $v_{n,m}(t)=w_{P,n,m}+w_{I,n,m}\tilde{x}_n(t)$, the signal received at the single-antenna receiver due to transmit antenna $m$ can be expressed as the sum of two contributions, namely one originating from WPT $y_{P,m}(t)$ and the other from WIT $y_{I,m}(t)$, namely
\begin{align}
\!y_{m}(t)\!&=y_{P,m}(t)+y_{I,m}(t),\nonumber\\
&=\!\Re\left\{\sum_{l=0}^{L-1}\sum_{n=0}^{N-1} \alpha_l v_{n,m}(t-\tau_l) e^{j 2\pi f_n (t-\tau_l)+\zeta_{n,m,l}}\right\}\!,\nonumber\\
&\approx\!\Re\left\{\sum_{n=0}^{N-1} h_{n,m}\left(w_{P,n,m}\!+\!w_{I,n,m}\tilde{x}_n(t)\right) e^{j 2\pi f_n t}\right\}\label{received_signal_ant_m}
\end{align}
where we have assumed $\max_{l\neq l'}\left|\tau_l-\tau_{l'}\right|<<1/B_s$ so that $v_{n,m}(t)$ and $\tilde{x}_n(t)$ for each subband are narrowband signals, thus $v_{n,m}(t-\tau_l)=v_{n,m}(t)$ and $\tilde{x}_n(t-\tau_l)=\tilde{x}_n(t)$, $\forall l$. The quantity $h_{n,m}=\!A_{n,m}e^{j \bar{\psi}_{n,m}}\!=\!\sum_{l=0}^{L-1}\alpha_l e^{j(-2\pi f_n\tau_l+\zeta_{n,m,l})}$ is the channel frequency response between antenna $m$ and the receive antenna at frequency $f_n$. 

\par Stacking up all transmit signals across all antennas, we can write the transmit WPT and WIT signal vectors as 
\begin{align}
\mathbf{x}_{P}(t)&=\Re\left\{\sum_{n=0}^{N-1}\mathbf{w}_{P,n}e^{j 2\pi f_n t}\right\},\\
\mathbf{x}_{I}(t)&=\Re\left\{\sum_{n=0}^{N-1}\mathbf{w}_{I,n} \tilde{x}_n(t) e^{j 2\pi f_n t}\right\}
\end{align}
where $\mathbf{w}_{P/I,n}=\big[\begin{array}{ccc}w_{P/I,n,1}&\ldots & w_{P/I,n,M}\end{array}\big]^T$. Similarly, we define the vector channel as $\mathbf{h}_n\!=\!\big[\begin{array}{ccc}h_{n,1}\!&\! ... \!&\! h_{n,M}\end{array}\big]$. The total received signal comprises the sum of \eqref{received_signal_ant_m} over all transmit antennas, namely
\begin{align}
y(t)&=y_{P}(t)+y_{I}(t),\nonumber\\
&=\Re\left\{\sum_{n=0}^{N-1}\mathbf{h}_n\left(\mathbf{w}_{P,n}+\mathbf{w}_{I,n} \tilde{x}_{n}\right) e^{j 2\pi f_n t}\right\}.\label{y_t}
\end{align}

\par The magnitudes and phases of the sinewaves can be collected into $N \times M$ matrices $\mathbf{S}_P$ and $\mathbf{\Phi}_P$. The $(n,m)$ entry of $\mathbf{S}_P$ and $\mathbf{\Phi}_P$ write as $s_{P,n,m}$ and $\phi_{P,n,m}$, respectively. Similarly, we define $N \times M$ matrices such that the $(n,m)$ entry of matrix  $\mathbf{S}_I$ and $\mathbf{\Phi}_I$ write as $s_{I,n,m}$ and $\phi_{I,n,m}$, respectively. We define the average power of the WPT and WIT waveforms as $P_P=\frac{1}{2}\left\|\mathbf{S}_P\right\|_F^2$ and $P_I=\frac{1}{2}\left\|\mathbf{S}_I\right\|_F^2$. Due to the superposition of the two waveforms, the total average transmit power constraint writes as $P_P+P_I\leq P$.

\par Following Fig \ref{WIPT_transceiver}(b), using a power splitter with a power splitting ratio $\rho$ and assuming perfect matching (as in Section \ref{antenna_eq_circuit}), the input voltage signals $\sqrt{\rho R_{ant}}y(t)$ and $\sqrt{(1-\rho)R_{ant}}y(t)$ are respectively conveyed to the energy harvester (EH) and the information decoder (ID). 

\begin{remark}\label{twofold_benefit}
As it will appear clearer throughout the paper, the benefit of choosing a deterministic multisine power waveform over other types of power waveform (e.g. modulated, pseudo-random) is twofold: 1) \textit{energy benefit}: multisine will be shown to be superior to a modulated waveform, 2) \textit{rate benefit}: multisine is deterministic and therefore does not induce any rate loss at the communication receiver.
\end{remark}

\begin{remark}\label{remark_distribution} It is worth noting the effect of the deterministic multisine waveform on the input distribution in \eqref{SWIPT_WF}. Recall that $x_{n,m}=\tilde{s}_{I,n,m}e^{j \tilde{\phi}_{I,n,m}} \sim \mathcal{CN}(0,s_{I,n,m}^2)$. Hence $w_{P,n,m}+x_{n,m}\sim \mathcal{CN}(w_{P,n,m},s_{I,n,m}^2)$ and the effective input distribution on a given frequency and antenna is not zero mean\footnote{If using OFDM, $x_{I,m}(t)$ and $x_{m}(t)$ are OFDM waveforms with CSCG inputs and non-zero mean Gaussian inputs, respectively.}. The magnitude $\left|w_{P,n,m}+x_{n,m}\right|$ is Ricean distributed with a K-factor on frequency $n$ and antenna $m$ given by $K_{n,m}=s_{P,n,m}^2/s_{I,n,m}^2$.   
\end{remark}

\begin{remark}
The superposition of information and power signals has appeared in other works, but for completely different purposes; namely for multiuser WIPT in \cite{Xu:2014}, collaborative WIPT in interference channel in \cite{SLee:2015,HLee:2015}, and for secrecy reasons in \cite{Liu:2014,Ng:2014}. Since those works relied on the linear model, the superposition was not motivated by the rectifier nonlinearity.
Moreover, the properties of the power signals are completely different. While the power signal is a deterministic multisine waveform leading to non-zero mean Gaussian input and the twofold benefit (Remark \ref{twofold_benefit}) in this work, it is complex (pseudo-random) Gaussian $\sim\mathcal{CN}(\mathbf{0},\mathbf{\Sigma})$ in those works.
\end{remark}

\vspace{-0.2cm}
\subsection{Information Decoder} 
\par Since $x_{P,m}(t)$ does not contain any information, it is deterministic. This has for consequence that the differential entropy of $v_{n,m}$ and $x_{n,m}$ are identical (because translation does not change the differential entropy) and the achievable rate is always equal to 
\begin{equation}
I(\mathbf{S}_I,\mathbf{\Phi}_I,\rho)\!=\!\sum_{n=0}^{N-1} \log_2\left(1+\frac{(1-\rho)\left|\mathbf{h}_n\mathbf{w}_{I,n}\right|^2}{\sigma_n^2} \right),\label{R}
\end{equation}
where $\sigma_n^2$ is the variance of the AWGN from the antenna and the RF-to-baseband down-conversion on tone $n$.


\par Naturally, $I(\mathbf{S}_I,\mathbf{\Phi}_I,\rho)$ is larger than the maximum rate achievable when $\rho=0$, i.e. $I(\mathbf{S}_I^{\star},\mathbf{\Phi}_I^{\star},0)$, which is obtained by performing Maximum Ratio Transmission (MRT) on each subband and water-filling power allocation across subbands.

\par The rate \eqref{R} is achievable irrespectively of the receiver architecture, e.g.\ with and without waveform cancellation. In the former case, after down-conversion from RF-to-baseband (BB) and ADC, the contribution of the power waveform is subtracted from the received signal (as illustrated in Fig \ref{WIPT_transceiver}(b))\footnote{If using OFDM, conventional OFDM processing (removing the cyclic prefix and performing FFT) is then conducted in the BB receiver.}. In the latter case, the ``Power WF cancellation'' box of Fig \ref{WIPT_transceiver}(b) is removed and the BB receiver decodes the translated version of the codewords.

\subsection{Energy Harvester}\label{section_EH}
In \cite{Clerckx:2016b}, a tractable model of the rectifier nonlinearity in the presence of multi-carrier unmodulated (deterministic multisine) excitation was derived and its validity verified through circuit simulations. In this paper, we reuse the same model and further expand it to modulated excitation. The randomness due to information symbols $\tilde{x}_n$ impacts the amount of harvested energy and needs to be captured in the model.   

\begin{figure}
   \centerline{\includegraphics[width=0.9\columnwidth]{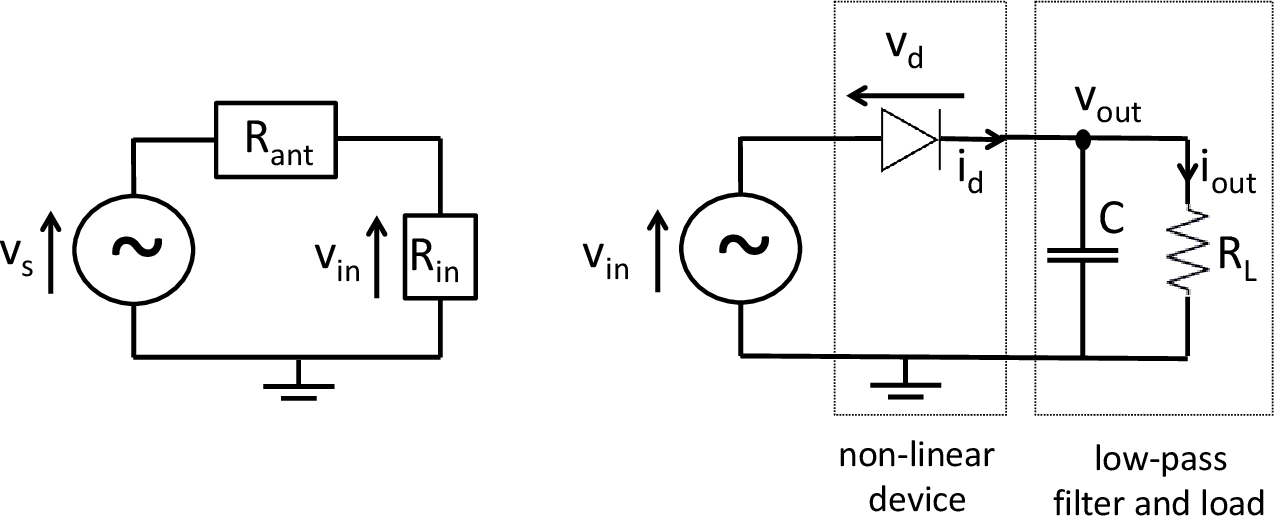}}
  \caption{Antenna equivalent circuit (left) and a single diode rectifier (right).}
  \label{antenna_model}
	\vspace{-0.2cm}
\end{figure}

\subsubsection{Antenna and Rectifier}\label{antenna_eq_circuit}
The signal impinging on the antenna is $y(t)$ and has an average power $P_{av}=\mathcal{E}\big\{\left|y(t)\right|^2\big\}$. 
A lossless antenna is modelled as a voltage source $v_s(t)$ followed by a
series resistance\footnote{Assumed real for simplicity. A more general model can be found in \cite{Clerckx:2017}.} $R_{ant}$ (Fig \ref{antenna_model} left). Let $Z_{in} \!=\! R_{in} + j X_{in}$ denote the
input impedance of the rectifier with the matching network.
Assuming perfect matching ($R_{in} = R_{ant}$, $X_{in} = 0$), due to the power splitter, a fraction $\rho$ of the
available RF power $P_{av}$ is transferred to the rectifier and
absorbed by $R_{in}$, so that the actual input power to the rectifier is $P_{in} = \rho\mathcal{E}\big\{\left|y(t)\right|^2\big\}\!=\!\mathcal{E}\big\{\left|v_{in}(t)\right|^2\big\}/R_{in}$ and $v_{in}(t)\!=\!v_{s}(t)/2$. Hence, $v_{in}(t)$ can be formed as $v_{in}(t)\!=\!y(t)\sqrt{\rho R_{in}}\!=\!y(t)\sqrt{\rho R_{ant}}$.
We also assume that the antenna noise is too small to be harvested.

\par Let us now look at Fig \ref{antenna_model}(right) and consider a rectifier composed of a single series diode\footnote{The model holds also for more general rectifiers as shown in \cite{Clerckx:2017}.} followed by a low-pass filter with load. Denoting the voltage drop across the diode as $v_d(t)=v_{in}(t)-v_{out}(t)$ where $v_{in}(t)$ is the input voltage to the diode and $v_{out}(t)$ is the output voltage across the load resistor, a tractable behavioral diode model is obtained by Taylor series expansion of the diode characteristic equation $i_d(t)=i_s \big(e^{\frac{v_d(t)}{n v_t}}-1 \big)$ (with $i_s$ the reverse bias saturation current, $v_t$ the thermal voltage, $n$ the ideality factor assumed equal to $1.05$) around a quiescent operating point $v_d=a$, namely $i_d(t)=\sum_{i=0}^{\infty }k_i' \left(v_d(t)-a\right)^i$
where $k_0'=i_s\big(e^{\frac{a}{n v_t}}-1\big)$ and $k_i'=i_s\frac{e^{\frac{a}{n v_t}}}{i!\left(n v_t\right)^i}$, $i=1,\ldots,\infty$.
Assume a steady-state response and an ideal low pass filter such that $v_{out}(t)$ is at constant DC level. Choosing $a=\mathcal{E} \left\{ v_d(t) \right\}=-v_{out}$, we can write $i_d(t)=\sum_{i=0}^{\infty }k_i' v_{in}(t)^i=\sum_{i=0}^{\infty }k_i' \rho^{i/2} R_{ant}^{i/2} y(t)^i$.

\par Under the ideal rectifier assumption and a deterministic incoming waveform $y(t)$, the current delivered to the load in a steady-state response is constant and given by $i_{out}=\mathcal{A}\left\{i_d(t)\right\}$. In order to make the optimization tractable, we truncate the Taylor expansion to the $n_o^{th}$ order. A nonlinear model truncates the Taylor expansion to the $n_o^{th}$ order but retains the fundamental nonlinear behavior of the diode while a linear model truncates to the second order term.

\subsubsection{Linear and Nonlinear Models}
After truncation, the output DC current approximates as 
\begin{equation}\label{diode_model}
i_{out}=\mathcal{A}\left\{i_d(t)\right\}\approx \sum_{i=0}^{n_o} k_i' \rho^{i/2} R_{ant}^{i/2} \mathcal{A}\left\{y(t)^i\right\}.
\end{equation}
\par Let us first consider a \textit{multi-carrier unmodulated (multisine) waveform}, i.e.\ $y(t)=y_P(t)$. Following \cite{Clerckx:2016b}, we get an approximation of the DC component of the current at the output of the rectifier (and the low-pass filter) with a multisine excitation over a multipath channel as 
\begin{equation}\label{diode_model_2}
i_{out}\approx k_0'+ \sum_{i \hspace{0.1cm}\textnormal{even}, i\geq 2}^{n_o} k_i' \rho^{i/2} R_{ant}^{i/2} \mathcal{A}\left\{y_P(t)^i\right\} 
\end{equation}
where $\mathcal{A}\left\{y_P(t)^2\right\}$ and $\mathcal{A}\left\{y_P(t)^4\right\}$ are detailed in \eqref{y_DC_2_2} and \eqref{y_DC_4_3}, respectively (at the top of next page). The linear model is a special case of the nonlinear model and is obtained by truncating the Taylor expansion to order 2 ($n_o=2$).  
\begin{table*}
\begin{align}
\mathcal{A}\left\{y_P(t)^2\right\}&=\frac{1}{2}\left[\sum_{n=0}^{N-1} \left|\mathbf{h}_n\mathbf{w}_{P,n} \right|^2\right]=\frac{1}{2}\left[\sum_{n=0}^{N-1} \sum_{m_0,m_1} s_{P,n,m_0}s_{P,n,m_1}A_{n,m_0}A_{n,m_1}\cos\left(\psi_{P,n,m_0}-\psi_{P,n,m_1}\right)\right],\label{y_DC_2_2}\\
\mathcal{A}\left\{y_P(t)^4\right\}
&=\frac{3}{8}\Re\left\{\sum_{\mycom{n_0,n_1,n_2,n_3}{n_0+n_1=n_2+n_3}}\mathbf{h}_{n_0}\mathbf{w}_{P,n_0}\mathbf{h}_{n_1}\mathbf{w}_{P,n_1}\left(\mathbf{h}_{n_2}\mathbf{w}_{P,n_2}\right)^*\left(\mathbf{h}_{n_3}\mathbf{w}_{P,n_3}\right)^*\right\},\label{y_DC_4_1b}\\
&=\frac{3}{8}\left[\sum_{\mycom{n_0,n_1,n_2,n_3}{n_0+n_1=n_2+n_3}}\sum_{\mycom{m_0,m_1,}{m_2,m_3}}\Bigg[\prod_{j=0}^3s_{P,n_j,m_j}A_{n_j,m_j}\Bigg]\cos(\psi_{P,n_0,m_0}+\psi_{P,n_1,m_1}-\psi_{P,n_2,m_2}-\psi_{P,n_3,m_3})\right].\label{y_DC_4_3}
\end{align}\hrulefill
\end{table*}

\par Let us then consider the \textit{multi-carrier modulated waveform}, i.e.\ $y(t)=y_I(t)$. It can be viewed as a multisine waveform for a fixed set of input symbols $\left\{\tilde{x}_n\right\}$. Hence, we can also write the DC component of the current at the output of the rectifier (and the low-pass filter) with a multi-carrier modulated excitation and fixed set of input symbols over a multipath channel as $k_0'+ \sum_{i \hspace{0.1cm}\textnormal{even}, i\geq 2}^{n_o} k_i' \rho^{i/2} R_{ant}^{i/2} \mathcal{A}\left\{y_I(t)^i\right\}$.
Similar expressions as \eqref{y_DC_2_2} and \eqref{y_DC_4_3} can be written for $\mathcal{A}\left\{y_I(t)^2\right\}$ and $\mathcal{A}\left\{y_I(t)^4\right\}$ for a fixed set of input symbols $\left\{\tilde{x}_n\right\}$. However, contrary to the multisine waveform, the input symbols $\left\{\tilde{x}_n\right\}$ of the modulated waveform change randomly at symbol rate $1/B_s$. For a given channel impulse response, the proposed model for the DC current with a modulated waveform is obtained as  
\begin{equation}\label{diode_model_2_OFDM}
i_{out}\approx k_0'+ \sum_{i \hspace{0.1cm}\textnormal{even}, i\geq 2}^{n_o} k_i' \rho^{i/2} R_{ant}^{i/2} \mathcal{E}_{\left\{\tilde{x}_n\right\}}\left\{\mathcal{A}\left\{y_I(t)^i\right\}\right\},
\end{equation}
by taking the expectation over the distribution of the input symbols $\left\{\tilde{x}_n\right\}$.
For $\mathcal{E}\left\{\mathcal{A}\left\{y_I(t)^i\right\}\right\}$ with $i$ even, the DC component is first extracted for a given set of amplitudes $\left\{\tilde{s}_{I,n,m}\right\}$ and phases $\big\{\tilde{\phi}_{I,n,m}\big\}$ and then expectation is taken over the randomness of the input symbols $\tilde{x}_n$. Due to the i.i.d.\ CSCG distribution of the input symbols, $\left|\tilde{x}_n\right|^2$ is exponentially distributed with $\mathcal{E}\left\{\big|\tilde{x}_n\right|^2\big\}=1$ and $\phi_{\tilde{x}_n}$ is uniformly distributed. From the moments of an exponential distribution, we also have that $\mathcal{E}\left\{\big|\tilde{x}_n\right|^4\big\}=2$. We can then express \eqref{E_y_I_2} and \eqref{E_y_I_40} as a function of $s_{I,n,m}$ and $\psi_{I,n,m}=\phi_{I,n,m}+\bar{\psi}_{n,m}$. Note that this factor of $\mathcal{E}\left\{\big|\tilde{x}_n\right|^4\big\}=2$ does not appear in \eqref{y_DC_4_3} due to the absence of modulation, which explains why \eqref{y_DC_4_3} and \eqref{E_y_I_40} enjoy a multiplicative factor of $\frac{3}{8}$ and $\frac{6}{8}$, respectively. Here again, the linear model is obtained by truncating to $n_o=2$.
\begin{table*}
\begin{align} 
\mathcal{E}\left\{\mathcal{A}\left\{y_I(t)^2\right\}\right\}
&=\frac{1}{2}\left[\sum_{n=0}^{N-1} \sum_{m_0,m_1} s_{I,n,m_0}s_{I,n,m_1}A_{n,m_0}A_{n,m_1}\cos\left(\psi_{I,n,m_0}-\psi_{I,n,m_1}\right)\right]= \frac{1}{2}\left[\sum_{n=0}^{N-1} \left|\mathbf{h}_n\mathbf{w}_{I,n} \right|^2 \right]\label{E_y_I_2}\\
\mathcal{E}\left\{\mathcal{A}\left\{y_I(t)^4\right\}\right\}
&=\frac{6}{8}\Bigg[\sum_{n_0,n_1}\sum_{\mycom{m_0,m_1,}{m_2,m_3}}\Bigg[\prod_{j=0,2}s_{I,n_0,m_j}A_{n_0,m_j}\Bigg]\Bigg[\prod_{j=1,3}s_{I,n_1,m_j}A_{n_1,m_j}\Bigg]
\cos(\psi_{I,n_0,m_0}\!+\!\psi_{I,n_1,m_1}\!-\!\psi_{I,n_0,m_2}\!-\!\psi_{I,n_1,m_3})\Bigg]\label{E_y_I_40}\\
&= \frac{6}{8}\left[\sum_{n=0}^{N-1} \left|\mathbf{h}_n\mathbf{w}_{I,n} \right|^2 \right]^2\label{E_y_I_4}
\end{align}
\hrulefill
\end{table*} 
 
\par Let us finally consider the \textit{superposed waveform}, i.e.\ $y(t)=y_{P}(t)+y_{I}(t)$. Both $y_{P}(t)$ and $y_{I}(t)$ waveforms now contribute to the DC component
\begin{equation}
i_{out}\approx k_0'+\sum_{i \hspace{0.1cm}\textnormal{even}, i\geq 2}^{n_o} k_i' \rho^{i/2} R_{ant}^{i/2} \mathcal{E}_{\left\{\tilde{x}_n\right\}}\big\{\mathcal{A}\big\{y(t)^i\big\}\big\}.
\end{equation}
Taking for instance $n_o=4$ and further expanding the term $\mathcal{E}_{\left\{\tilde{x}_n\right\}}\big\{\mathcal{A}\big\{y(t)^i\big\}\big\}$ using the fact that $\mathcal{E}\left\{\mathcal{A}\left\{y_P(t)y_I(t)\right\}\right\}=0$, $\mathcal{E}\left\{\mathcal{A}\left\{y_P(t)^3y_I(t)\right\}\right\}=0$, $\mathcal{E}\left\{\mathcal{A}\left\{y_P(t)y_I(t)^3\right\}\right\}=0$ and $\mathcal{E}\left\{\mathcal{A}\left\{y_P(t)^2y_I(t)^2\right\}\right\}=\mathcal{A}\left\{y_P(t)^2\right\}\mathcal{E}\left\{\mathcal{A}\left\{y_I(t)^2\right\}\right\}$, $i_{out}$ can be written as
\begin{multline}
i_{out}\approx k_0'+k_2' \rho R_{ant}\mathcal{A}\left\{y_P(t)^2\right\}+k_4'\rho^2 R_{ant}^2\mathcal{A}\left\{y_P(t)^4\right\}\\
+k_2' \rho R_{ant}\mathcal{E}\left\{\mathcal{A}\left\{y_I(t)^2\right\}\right\}+k_4'\rho^2 R_{ant}^2 \mathcal{E}\left\{\mathcal{A}\left\{y_I(t)^4\right\}\right\}\\
+6k_4'\rho^2 R_{ant}^2 \mathcal{A}\left\{y_P(t)^2\right\}\mathcal{E}\left\{\mathcal{A}\left\{y_I(t)^2\right\}\right\}.\label{i_out_no4}
\end{multline}

\begin{observation}\label{obs_main} The \textit{linear model} highlights that there is no difference in using a multi-carrier unmodulated (multisine) waveform and a multi-carrier modulated (e.g.\ OFDM) waveform for WPT, since according to this model the harvested energy is a function of $\sum_{n=0}^{N-1} \left|\mathbf{h}_n\mathbf{w}_{P/I,n} \right|^2$, as seen from \eqref{y_DC_2_2} and \eqref{E_y_I_2}. Hence modulated and unmodulated waveforms are equally suitable. On the other hand, the \textit{nonlinear model} highlights that there is a clear difference between using a multi-carrier unmodulated over a multi-carrier modulated waveform in WPT. Indeed, from \eqref{E_y_I_2} and \eqref{E_y_I_4} of the modulated waveform, both the second and fourth order terms exhibit the same behavior and same dependencies, namely they are both exclusively function of $\sum_{n=0}^{N-1} \left|\mathbf{h}_n\mathbf{w}_{I,n} \right|^2$. That suggests that for a multi-carrier modulated waveform with CSCG inputs, the linear and nonlinear models are equivalent, i.e.\ there is no need in modeling the fourth and higher order term. On the other hand, for the unmodulated waveform, the second and fourth order terms, namely \eqref{y_DC_2_2} and \eqref{y_DC_4_3}, exhibit clearly different behaviors with the second order term being linear and the fourth order being nonlinear and function of terms expressed as the product of contributions from different frequencies.
\end{observation}

\begin{remark} The linear model is motivated by its simplicity rather than its accuracy and is the popular model used throughout the WIPT literature, e.g. \cite{Zhang:2013}. Indeed, it is always assumed that the harvested DC power is modeled as $\eta P_{in}(y(t))=\eta \mathcal{E}\left\{\mathcal{A}\left\{y(t)^2\right\}\right\}$ where $\eta$ is the RF-to-DC conversion efficiency assumed constant. By assuming $\eta$ constant, those works effectively only care about maximizing the input power $P_{in}(y(t))$ (function of $y(t)$) to the rectifier, i.e.\ the second order term (or linear term) $\mathcal{E}\left\{\mathcal{A}\left\{y(t)^2\right\}\right\}$ in the Taylor expansion. Unfortunately this is inaccurate as $\eta$ is not a constant and is itself a function of the input waveform (power and shape) to the rectifier, as recently highlighted in the communication literature \cite{Clerckx:2015,Clerckx:2016b,Zeng:2017,Boshkovska:2015} but well recognized in the RF literature \cite{OptBehaviour,Costanzo:2016}. This linear model was shown through circuit simulations in \cite{Clerckx:2016b} to be inefficient to design multisine waveform but also inaccurate to predict the behavior of such waveforms in the practical low-power regime (-30dBm to 0dBm). On the other hand, the nonlinear model, rather than explicitly expressing the DC output power as $\eta(y(t)) P_{in}(y(t))$ with $\eta(y(t))$ a function of the input signal power and shape, it directly expresses the output DC current as a function of $y(t)$ (and therefore as a function of the transmit signal and wireless channel) and leads to a more tractable formulation. Such a nonlinear model with $n_o=4$ has been validated for the design of multisine waveform in \cite{Clerckx:2015,Clerckx:2016b,Clerckx:2017} using circuit simulators with various rectifier topologies and input power and in \cite{Kim:2017} through prototyping and experimentation. Nevertheless, the use of a linear vs a nonlinear model for the design of WPT based on other types of waveforms and the design of WIPT has never been addressed so far. 
\end{remark}

\begin{remark}\label{EH_remark} 
The above model deals with the \textit{diode nonlinearity} under ideal low pass filter and perfect impedance matching. However there exist other sources of nonlinearities in a rectifier, e.g.\ impedance mismatch, breakdown voltage and harmonics. Recently, another nonlinear model has emerged in \cite{Boshkovska:2015}. This model accounts for the fact that for a given rectifier design, the RF-to-DC conversion efficiency $\eta$ is a function of the input power and sharply decreases once the input power has reached the diode breakdown region. This leads to a \textit{saturation nonlinearity} where the output DC power saturates beyond a certain input power level. There are multiple differences between those two models.\\
\textit{First}, our diode nonlinearity model assumes the rectifier is not operating in the diode breakdown region. Circuit evaluations in \cite{Clerckx:2016b,Clerckx:2017} and in Section \ref{eval_circuit} also confirm that the rectifier never reached the diode breakdown voltage under all investigated scenarios. We therefore do not model the saturation effect. On the other hand, \cite{Boshkovska:2015} assumes the rectifier can operate in the breakdown region and therefore models the saturation. However, it is to be reminded that operating diodes in the breakdown region is not the purpose of a rectifier and should be avoided. A rectifier is designed in such a way that current flows in only one direction, not in both directions as it would occur in the breakdown region. Hence, \cite{Boshkovska:2015} models a saturation nonlinearity effect that occurs in an operating region where one does not wish to operate in. In other words, the rectifier is pushed in an input power range quite off from the one it has originally been designed for. This may only occur in applications where there is little guarantee to operate the designed rectifier below that breakdown edge. \\
\textit{Second}, the diode nonlinearity is a fundamental, unavoidable and intrinsic property of any rectifier, i.e. any rectifier, irrespectively of its design, topology or implementation, is always made of a nonlinear device (most commonly Schottky diode) followed by a low pass filter with load. This has for consequence that the diode nonlinearity model is general and valid for a wide range of rectifier design and topology (with one and multiple diodes) as shown in \cite{Clerckx:2017}. Moreover, since it is driven by the physics of the rectenna, it analytically links the output DC metric to the input signal through the diode I-V characteristics. On the other hand, the saturation nonlinearity in \cite{Boshkovska:2015} is circuit-specific and modeled via curve fitting based on measured data. Hence changing the diode or the rectifier topology would lead to a different behavior. More importantly, the saturation effect, and therefore the corresponding nonlinearity, is actually avoidable by properly designing the rectifier for the input power range of interest. A common strategy is to use an adaptive rectifier whose configuration changes as a function of the input power level, e.g.\ using a single-diode rectifier at low input power and multiple diodes rectifier at higher power, so as to generate consistent and non-vanishing $\eta$ over a significantly extended operating input power range \cite{Sun:2013,Li:2014}.\\
\textit{Third}, the diode nonlinearity model accomodates a wide range of multi-carrier modulated and unmodulated input signals and is therefore a function of the input signal power, shape and modulation. The saturation nonlinearity model in \cite{Boshkovska:2015} is restricted to a continuous wave input signal and is a function of its power. Hence it does not reflect the dependence of the output DC power to modulation and waveform designs.  \\
\textit{Fourth}, the diode nonlinearity is a beneficial feature that is to be exploited as part of the waveform design to boost the output DC power, as shown in \cite{Clerckx:2016b}. The saturation nonlinearity is detrimental to performance and should therefore be avoided by operating in the non-breakdown region and using properly designed rectifier for the input power range of interest.\\
\textit{Fifth}, the diode nonlinearity is more meaningful in the low-power regime (-30dBm to 0dBm with state-of-the-art rectifiers\footnote{At lower power levels, the diode may not turn on.}) while the saturation nonlinearity is relevant in the high power regime (beyond 0dBm input power).
\end{remark} 

\vspace{-0.2cm}
\section{WIPT Waveform Optimization and Rate-Energy Region Characterization}\label{section_SWIPT_waveform}
Leveraging the energy harvester model, we now aim at characterizing the rate-energy region of the proposed WIPT architecture. We define the achievable rate-energy region as
\begin{multline}
C_{R-I_{DC}}(P)\triangleq\Big\{(R,I_{DC}):R\leq I,\Big. \\
\Big.I_{DC}\leq i_{out}, \frac{1}{2}\big[\left\|\mathbf{S}_I\right\|_F^2+\left\|\mathbf{S}_P\right\|_F^2\big]\leq P \Big\}.
\end{multline}
Assuming the CSI (in the form of frequency response $h_{n,m}$) is known to the transmitter, we aim at finding the optimal values of amplitudes, phases and power splitting ratio, denoted as  $\mathbf{S}_P^{\star}$,$\mathbf{S}_I^{\star}$,$\mathbf{\Phi}_P^{\star}$,$\mathbf{\Phi}_I^{\star},\rho^{\star}$, so as to enlarge as much as possible the rate-energy region. We derive a methodology that is general to cope with any truncation order $n_o$\footnote{We display terms for $n_o\!\leq \! 4$ but the derived algorithm works for any $n_o$.}.

\par Characterizing such a region involves solving the problem
\begin{align}\label{P1}
\max_{\mathbf{S}_P,\mathbf{S}_I,\mathbf{\Phi}_P,\mathbf{\Phi}_I,\rho} \hspace{0.2cm} &i_{out}(\mathbf{S}_P,\mathbf{S}_I,\mathbf{\Phi}_P,\mathbf{\Phi}_I,\rho) \\
\textnormal{subject to} \hspace{0.3cm} &\frac{1}{2}\big[\left\|\mathbf{S}_I\right\|_F^2+\left\|\mathbf{S}_P\right\|_F^2\big]\leq P.\label{P1_2}
\end{align}
Following \cite{Clerckx:2016b}, \eqref{P1}-\eqref{P1_2} can equivalently be written as
\begin{align}\label{P1_eq}
\max_{\mathbf{S}_P,\mathbf{S}_I,\mathbf{\Phi}_P,\mathbf{\Phi}_I,\rho} \hspace{0.2cm} &z_{DC}(\mathbf{S}_P,\mathbf{S}_I,\mathbf{\Phi}_P,\mathbf{\Phi}_I,\rho) \\
\textnormal{subject to} \hspace{0.3cm} &\frac{1}{2}\big[\left\|\mathbf{S}_I\right\|_F^2+\left\|\mathbf{S}_P\right\|_F^2\big]\leq P,\label{P1_eq_2}
\end{align}
with $z_{DC}\!=\!\sum_{\mycom{i \hspace{0.1cm}\textnormal{even},}{i\geq 2}}^{n_o}\! k_i \rho^{i/2} R_{ant}^{i/2} \mathcal{E}_{\left\{\tilde{x}_n\right\}}\!\left\{\!\mathcal{A}\left\{y(t)^i\right\}\!\right\}$
where we define $k_i\!=\!\frac{i_s}{i!\left(n v_t\right)^i}$. Assuming $i_s\!=\!5 \mu A$, a diode ideality factor $n\!=\!1.05$ and $v_t\!=\!25.86 mV$, we get $k_2\!=\!0.0034$ and $k_4\!=\!0.3829$. For $n_o\!=\!4$, similarly to \eqref{i_out_no4}, we can compute $z_{DC}$ as in \eqref{z_DC_SWIPT}.
\begin{table*}
\begin{multline}\label{z_DC_SWIPT}
z_{DC}(\mathbf{S}_P,\mathbf{S}_I,\mathbf{\Phi}_P,\mathbf{\Phi}_I,\rho)=k_2 \rho R_{ant}\mathcal{A}\left\{y_P(t)^2\right\}+k_4\rho^2 R_{ant}^2\mathcal{A}\left\{y_P(t)^4\right\}+k_2 \rho R_{ant}\mathcal{E}\left\{\mathcal{A}\left\{y_I(t)^2\right\}\right\}\\
+k_4\rho^2 R_{ant}^2 \mathcal{E}\left\{\mathcal{A}\left\{y_I(t)^4\right\}\right\}+6k_4\rho^2 R_{ant}^2 \mathcal{A}\left\{y_P(t)^2\right\}\mathcal{E}\left\{\mathcal{A}\left\{y_I(t)^2\right\}\right\}.
\end{multline} 
\hrulefill
\end{table*}

\par This enables to re-define the achievable rate-energy region in terms of $z_{DC}$ rather than $i_{out}$ as follows
\begin{multline}\label{R_zDC_region}
C_{R-I_{DC}}(P)\triangleq\Big\{(R,I_{DC}):R\leq I,\Big. \\
\Big.I_{DC}\leq z_{DC}, \frac{1}{2}\big[\left\|\mathbf{S}_I\right\|_F^2+\left\|\mathbf{S}_P\right\|_F^2\big]\leq P \Big\},
\end{multline}
This definition of rate-energy region will be used in the sequel.

\vspace{-0.2cm}
\subsection{WPT-only: Energy Maximization}\label{OFDM_design}
In this section, we first look at energy maximization-only (with no consideration for rate) and therefore assume $\rho=1$. We study and compare the design of multi-carrier unmodulated (multisine) waveform ($y(t)=y_{P}(t)$) and modulated waveforms ($y(t)=y_{I}(t)$) under the linear and nonlinear models. Since a single waveform is transmitted (either unmodulated or modulated), the problem simply boils down to the following for $i\in\left\{P,I\right\}$
\begin{equation}\label{WPT_max}
\max_{\mathbf{S}_i,\mathbf{\Phi}_i} \hspace{0.2cm} z_{DC}(\mathbf{S}_i,\mathbf{\Phi}_i) \hspace{0.3cm} \textnormal{subject to} \hspace{0.3cm} \frac{1}{2}\left\|\mathbf{S}_i\right\|_F^2\leq P, 
\end{equation}
where $z_{DC}(\mathbf{S}_P,\!\mathbf{\Phi}_P)\!=\!\sum_{i \hspace{0.1cm}\textnormal{even}, i\geq 2}^{n_o}\! k_i R_{ant}^{i/2} \mathcal{A}\!\left\{y_P(t)^i\right\}$
for multi-carrier unmodulated (multisine) waveform, and $z_{DC}(\mathbf{S}_I,\!\mathbf{\Phi}_I)\!=\!\sum_{i \hspace{0.1cm}\textnormal{even}, i\geq 2}^{n_o}\! k_i R_{ant}^{i/2} \mathcal{E}_{\left\{\tilde{x}_n\right\}}\!\left\{\!\mathcal{A}\left\{y_I(t)^i\right\}\!\right\}$
for the multi-carrier modulated waveform. 

\par The problem of multisine waveform design with a linear and nonlinear rectenna model has been addressed in \cite{Clerckx:2016b}. The linear model leads to the equivalent problem $\max_{\mathbf{w}_{P,n}} \hspace{0.1cm} \sum_{n=0}^{N-1} \left|\mathbf{h}_n\mathbf{w}_{P,n} \right|^2$ subject to $\frac{1}{2}\sum_{n=0}^{N-1} \left\|\mathbf{w}_{P,n} \right\|^2\!\leq\! P$ whose solution is the adaptive single-sinewave (ASS) strategy
\begin{equation}\label{solution_2nd_order}
\mathbf{w}^{\star}_{P,n}=\left\{\begin{array}{l}
\sqrt{2P}\:\mathbf{h}_n^H/\left\|\mathbf{h}_n\right\|, \hspace{0.2cm} n=\bar{n}, \\
\mathbf{0}, \hspace{0.2cm} n\neq\bar{n}. 
\end{array}
\right.
\end{equation}
The ASS performs a matched (also called MRT) beamformer on a single sinewave, namely the one corresponding to the strongest channel $\bar{n}=\arg \max_{n}\left\|\mathbf{h}_n\right\|^2$. On the other hand, the nonlinear model leads to a posynomial maximization problem that can be formulated as a Reversed Geometric Program and solved iteratively. Interestingly, for multisine waveforms, the linear and nonlinear models lead to radically different strategies. The former favours transmission on a single frequency while the latter favours transmission over multiple frequencies. Design based on the linear model was shown to be inefficient and lead to significant loss over the nonlinear model-based design.

\par The design of multi-carrier modulated waveform is rather different. Recall that from \eqref{E_y_I_2} and \eqref{E_y_I_4}, both the second and fourth order terms are exclusively function of $\sum_{n=0}^{N-1} \left|\mathbf{h}_n\mathbf{w}_{I,n} \right|^2$. This shows that both the linear and nonlinear model-based designs of multi-carrier modulated waveforms for WPT lead to the ASS strategy and the optimum $\mathbf{w}^{\star}_{I,n}$ should be designed according to \eqref{solution_2nd_order}. This is in sharp contrast with the multisine waveform design and originates from the fact that the modulated waveform is subject to CSCG randomness due to the presence of input symbols $\tilde{x}_n$. Note that this ASS strategy has already appeared in the WIPT literature, e.g.\ in \cite{Huang:2013,Bayani:2015} with OFDM transmission. 

\vspace{-0.3cm}
\subsection{WIPT: A General Approach}\label{PC_ID}
\par We now aim at characterizing the rate-energy region of the proposed WIPT architecture. Looking at \eqref{R} and \eqref{z_DC_SWIPT}, it is easy to conclude that matched filtering w.r.t.\ the phases of the channel is optimal from both rate and harvested energy maximization perspective. This leads to the same phase decisions as for WPT in \cite{Clerckx:2015,Clerckx:2016b}, namely 
\begin{equation}\label{opt_phases}
\phi_{P,n,m}^{\star}=\phi_{I,n,m}^{\star}=-\bar{\psi}_{n,m}
\end{equation} 
and guarantees all arguments of the cosine functions in $\left\{\mathcal{A}\left\{y_P(t)^i\right\}\right\}_{i=2,4}$ (\eqref{y_DC_2_2}, \eqref{y_DC_4_3}) and in $\left\{\mathcal{E}\left\{\mathcal{A}\left\{y_I(t)^i\right\}\right\}\right\}_{i=2,4}$ (\eqref{E_y_I_2}, \eqref{E_y_I_40}) to be equal to 0. $\mathbf{\Phi}_P^{\star}$ and $\mathbf{\Phi}_I^{\star}$ are obtained by collecting $\phi_{P,n,m}^{\star}$ and $\phi_{I,n,m}^{\star}$ $\forall n,m$ into a matrix, respectively. 

\begin{table*}
\begin{align}\label{z_DC_SWIPT_final}
z_{DC}(\mathbf{S}_P,\mathbf{S}_I,\mathbf{\Phi}_P^{\star},\mathbf{\Phi}_I^{\star},\rho)&= \frac{k_2 \rho}{2}R_{ant}\left[\sum_{n=0}^{N-1} \sum_{m_0,m_1} \Bigg[\prod_{j=0}^1 s_{P,n,m_j}A_{n,m_j}\Bigg]\right]
+ \frac{3k_4\rho^2}{8}R_{ant}^2\left[\sum_{\mycom{n_0,n_1,n_2,n_3}{n_0+n_1=n_2+n_3}}\sum_{\mycom{m_0,m_1,}{m_2,m_3}}\Bigg[\prod_{j=0}^3s_{P,n_j,m_j}A_{n_j,m_j}\Bigg]\right]\nonumber\\
&\hspace{0.6cm}+\frac{k_2 \rho}{2}R_{ant}\left[\sum_{n=0}^{N-1} \sum_{m_0,m_1} \Bigg[\prod_{j=0}^1 s_{I,n,m_j}A_{n,m_j}\Bigg]\right]+ \frac{3 k_4\rho^2}{4}R_{ant}^2\left[\sum_{n=0}^{N-1} \sum_{m_0,m_1} \Bigg[\prod_{j=0}^1 s_{I,n,m_j}A_{n,m_j}\Bigg]\right]^2 \nonumber\\
&\hspace{0.6cm}+\frac{3k_4\rho^2}{2} R_{ant}^2\left[\sum_{n=0}^{N-1} \sum_{m_0,m_1} \Bigg[\prod_{j=0}^1 s_{P,n,m_j}A_{n,m_j}\Bigg]\right]\left[\sum_{n=0}^{N-1} \sum_{m_0,m_1} \Bigg[\prod_{j=0}^1 s_{I,n,m_j}A_{n,m_j}\Bigg]\right]
\end{align} 
\hrulefill
\end{table*}
\par With such phases $\mathbf{\Phi}_P^{\star}$ and $\mathbf{\Phi}_I^{\star}$, $z_{DC}(\mathbf{S}_P,\mathbf{S}_I,\mathbf{\Phi}_P^{\star},\mathbf{\Phi}_I^{\star},\rho)$ can be finally written as \eqref{z_DC_SWIPT_final}. Similarly we can write
\begin{equation}
I(\mathbf{S}_I,\mathbf{\Phi}_I^{\star},\rho)=\log_2\left(\prod_{n=0}^{N-1}\left(1+\frac{(1-\rho)}{\sigma_n^2} C_n\right)\right)\label{R_SWIPT_final} 
\end{equation}
where $C_n=\sum_{m_0,m_1} \prod_{j=0}^1 s_{I,n,m_j}A_{n,m_j}$. 

\par Recall from \cite{Chiang:2005} that a monomial is defined as the function $g:\mathbb{R}_{++}^{N}\rightarrow\mathbb{R}:g(\mathbf{x})=c x_1^{a_1}x_2^{a_2}\ldots x_N^{a_N}$
where $c>0$ and $a_i\in\mathbb{R}$. A sum of $K$ monomials is called a posynomial and can be written as $f(\mathbf{x})=\sum_{k=1}^K g_k(\mathbf{x})$ with $g_k(\mathbf{x})=c_k x_1^{a_{1k}}x_2^{a_{2k}}\ldots x_N^{a_{Nk}}$ where $c_k>0$. As we can see from \eqref{z_DC_SWIPT_final}, $z_{DC}(\mathbf{S}_P,\mathbf{S}_I,\mathbf{\Phi}_P^{\star},\mathbf{\Phi}_I^{\star},\rho)$ is a posynomial. 

\par In order to identify the achievable rate-energy region, we formulate the optimization problem as an energy maximization problem subject to transmit power and rate constraints
\begin{align}\label{SWIPT_opt_problem}
\max_{\mathbf{S}_P,\mathbf{S}_I,\rho} \hspace{0.3cm}&z_{DC}(\mathbf{S}_P,\mathbf{S}_I,\mathbf{\Phi}_P^{\star},\mathbf{\Phi}_I^{\star},\rho)\\
\textnormal{subject to} \hspace{0.3cm} &\frac{1}{2}\big[\left\|\mathbf{S}_I\right\|_F^2+\left\|\mathbf{S}_P\right\|_F^2\big]\leq P,\\
& I(\mathbf{S}_I,\mathbf{\Phi}_I^{\star},\rho)\geq \bar{R}.
\end{align}
It therefore consists in maximizing a posynomial subject to constraints. 
Unfortunately this problem is not a standard Geometric Program (GP) but it can be transformed to an equivalent problem by introducing an auxiliary variable $t_0$
\begin{align}\label{SWIPT_opt_problem_eq}
\min_{\mathbf{S}_P,\mathbf{S}_I,\rho,t_0} \hspace{0.3cm} &1/t_0\\
\textnormal{subject to} \hspace{0.3cm} &\frac{1}{2}\big[\left\|\mathbf{S}_I\right\|_F^2+\left\|\mathbf{S}_P\right\|_F^2\big]\leq P,\\
&t_0/z_{DC}(\mathbf{S}_P,\mathbf{S}_I,\mathbf{\Phi}_P^{\star},\mathbf{\Phi}_I^{\star},\rho)\leq1,\\
&2^{\bar{R}}/\left[\prod_{n=0}^{N-1}\left(1+\frac{(1-\rho)}{\sigma_n^2} C_n\right)\right]\leq 1.\label{SWIPT_opt_problem_eq_4}
\end{align}
This is known as a Reversed Geometric Program \cite{Chiang:2005}. A similar problem also appeared in the WPT waveform optimization \cite{Clerckx:2015}.
Note that $1/z_{DC}(\mathbf{S}_P,\mathbf{S}_I,\mathbf{\Phi}_P^{\star},\mathbf{\Phi}_I^{\star},\rho)$ and $1/\big[\prod_{n=0}^{N-1}\big(1+\frac{(1-\rho)}{\sigma_n^2} C_n\big)\big]$ are not posynomials, therefore preventing the use of standard GP tools. The idea is to replace the last two inequalities (in a conservative way) by making use of the arithmetic mean-geometric mean (AM-GM) inequality.

\par Let $\left\{g_k(\mathbf{S}_P,\mathbf{S}_I,\mathbf{\Phi}_P^{\star},\mathbf{\Phi}_I^{\star},\rho)\right\}$ be the monomial terms in the posynomial $z_{DC}(\mathbf{S}_P,\mathbf{S}_I,\mathbf{\Phi}_P^{\star},\mathbf{\Phi}_I^{\star},\rho)=\sum_{k=1}^K g_k(\mathbf{S}_P,\mathbf{S}_I,\mathbf{\Phi}_P^{\star},\mathbf{\Phi}_I^{\star},\rho)$. Similarly we define $\left\{g_{nk}(\mathbf{S}_I,\bar{\rho})\right\}$ as the set of monomials of the posynomial $1\!+\!\frac{\bar{\rho}}{\sigma_n^2}C_n\!=\!\sum_{k=1}^{K_n}g_{nk}(\mathbf{S}_I,\bar{\rho})$ with $\bar{\rho}\!=\!1\!-\!\rho$. For a given choice of $\left\{\gamma_k\right\}$ and $\left\{\gamma_{nk}\right\}$ with $\gamma_k,\gamma_{nk}\!\geq\! 0$ and $\sum_{k=1}^K \gamma_k\!=\!\sum_{k=1}^{K_n} \gamma_{nk}\!=\!1$, we perform single condensations and write the standard GP
\begin{align}\label{standard_GP_SWIPT}
\min_{\mathbf{S}_P,\mathbf{S}_I,\rho,\bar{\rho},t_0} \hspace{0.3cm} &1/t_0\\
\textnormal{subject to} \hspace{0.3cm} &\frac{1}{2}\big[\left\|\mathbf{S}_I\right\|_F^2+\left\|\mathbf{S}_P\right\|_F^2\big]\leq P,\\ 
&t_0\prod_{k=1}^K\left(\frac{g_k(\mathbf{S}_P,\mathbf{S}_I,\mathbf{\Phi}_P^{\star},\mathbf{\Phi}_I^{\star},\rho)}{\gamma_k}\right)^{-\gamma_k}\leq1,\\
&2^{\bar{R}}\prod_{n=0}^{N-1}\prod_{k=1}^{K_n}\left(\frac{g_{nk}(\mathbf{S}_I,\bar{\rho})}{\gamma_{nk}}\right)^{-\gamma_{nk}}\leq 1,\\
&\rho+\bar{\rho}\leq 1,\label{standard_GP_SWIPT_4}
\end{align}
that can be solved using existing software, e.g. CVX \cite{CVX}.
\par It is important to note that the choice of $\left\{\gamma_k,\gamma_{nk}\right\}$ plays a great role in the tightness of the AM-GM inequality. An iterative procedure can be used where at each iteration the standard GP \eqref{standard_GP_SWIPT}-\eqref{standard_GP_SWIPT_4} is solved for an updated set of $\left\{\gamma_k,\gamma_{nk}\right\}$. Assuming a feasible set of magnitude $\mathbf{S}_P^{(i-1)}$ and $\mathbf{S}_I^{(i-1)}$ and power splitting ratio $\rho^{(i-1)}$ at iteration $i-1$, compute at iteration $i$ $\gamma_k=\frac{g_k(\mathbf{S}_P^{(i-1)},\mathbf{S}_I^{(i-1)},\mathbf{\Phi}_P^{\star},\mathbf{\Phi}_I^{\star},\rho^{(i-1)})}{z_{DC}(\mathbf{S}_P^{(i-1)},\mathbf{S}_I^{(i-1)},\mathbf{\Phi}_P^{\star},\mathbf{\Phi}_I^{\star},\rho^{(i-1)})}$ $k=1,\ldots,K$ and $\gamma_{nk}= g_{nk}(\mathbf{S}_I^{(i-1)},\bar{\rho}^{(i-1)})/\big(1+\frac{\bar{\rho}^{(i-1)}}{\sigma_n^2}C_n(\mathbf{S}_I^{(i-1)})\big)$, $n=0,\ldots,N-1$, $k=1,\ldots,K_{n}$ and then solve problem \eqref{standard_GP_SWIPT}-\eqref{standard_GP_SWIPT_4} to obtain $\mathbf{S}_P^{(i)}$, $\mathbf{S}_I^{(i)}$ and $\rho^{(i)}$. Repeat the iterations till convergence. The whole optimization procedure is summarized in Algorithm \ref{Algthm_OPT_WIPT}.

\begin{algorithm}
\caption{WIPT Waveform and R-E Region}
\label{Algthm_OPT_WIPT}
\begin{algorithmic}[1]
\State \textbf{Initialize}: $i\gets 0$, $\bar{R}$, $\mathbf{\Phi}_P^{\star}$ and $\mathbf{\Phi}_I^{\star}$, $\mathbf{S}_P$, $\mathbf{S}_I$, $\rho$, $\bar{\rho}=1-\rho$, $z_{DC}^{(0)}=0$
\label{Algthm_OPT_WIPT_step_initialize}
\Repeat
    \State $i\gets i+1$, $\ddot{\mathbf{S}}_P\gets \mathbf{S}_P$, $\ddot{\mathbf{S}}_I\gets \mathbf{S}_I$, $\ddot{\rho}\gets \rho$, $\ddot{\bar{\rho}}\gets \bar{\rho}$
    \State $\gamma_k\gets g_k(\ddot{\mathbf{S}}_P,\ddot{\mathbf{S}}_I,\mathbf{\Phi}_P^{\star},\mathbf{\Phi}_I^{\star},\ddot{\rho})/z_{DC}(\ddot{\mathbf{S}}_P,\ddot{\mathbf{S}}_I,\mathbf{\Phi}_P^{\star},\mathbf{\Phi}_I^{\star},\ddot{\rho})$, $k=1,\ldots,K$  
    \label{Algthm_OPT_WIPT_step_gamma}
    \State $\gamma_{nk}\gets g_{nk}(\ddot{\mathbf{S}}_I,\ddot{\bar{\rho}})/\big(1+\frac{\ddot{\bar{\rho}}}{\sigma_n^2}C_n(\ddot{\mathbf{S}}_I)\big)$, $n=0,\ldots,N-1$, $k=1,\ldots,K_{n}$  
    \label{Algthm_OPT_WIPT_step_gamma_2}
    \State  $\mathbf{S}_P,\mathbf{S}_I,\rho,\bar{\rho} \gets \arg \min \eqref{standard_GP_SWIPT}-\eqref{standard_GP_SWIPT_4}$
    \label{Algthm_OPT_WIPT_step_OPT}
    \State $z_{DC}^{(i)} \gets z_{DC}(\mathbf{S}_P,\mathbf{S}_I,\mathbf{\Phi}_P^{\star},\mathbf{\Phi}_I^{\star},\rho)$
\Until{$\left|z_{DC}^{(i)} - z_{DC}^{(i-1)} \right| < \epsilon$ \text{or} $i=i_{\max}$ }
\end{algorithmic}
\end{algorithm}
 
Note that the successive approximation method used in the Algorithm 1 is also known as a sequential convex optimization or inner approximation method \cite{Marks:1978}. It cannot guarantee to converge to the global solution of the original problem, but only to yield a point fulfilling the KKT conditions \cite{Marks:1978,Chiang:2007}. 

\vspace{-0.2cm}
\subsection{WIPT: Decoupling Space and Frequency}\label{decoupling}
When $M>1$, previous section derives a general methodology to optimize the superposed waveform weights jointly across space and frequency. It is worth wondering whether we can decouple the design of the spatial and frequency domain weights without impacting performance. The optimal phases in \eqref{opt_phases} are those of a MRT beamformer. Looking at \eqref{R}, \eqref{y_DC_2_2}, \eqref{y_DC_4_1b}, \eqref{E_y_I_2} and \eqref{E_y_I_4}, the optimum weight vectors $\mathbf{w}_{P,n}$ and $\mathbf{w}_{I,n}$ that maximize the $2^{nd}$ and $4^{th}$ order terms and the rate, respectively, are MRT beamformers of the form
\begin{equation}\label{opt_weight}
\mathbf{w}_{P,n}=s_{P,n} \mathbf{h}_n^H/\left\|\mathbf{h}_n\right\|, \hspace{0.5cm} \mathbf{w}_{I,n}=s_{I,n} \mathbf{h}_n^H/\left\|\mathbf{h}_n\right\|,
\end{equation} 
such that, from \eqref{y_t}, $y_P(t)=\sum_{n=0}^{N-1}\left\|\mathbf{h}_n\right\| s_{P,n} \cos\left(w_n t\right)=\Re\big\{\sum_{n=0}^{N-1}\left\|\mathbf{h}_n\right\| s_{P,n} e^{j w_n t}\big\}$ and $y_I(t)=\sum_{n=0}^{N-1}\left\|\mathbf{h}_n\right\| s_{I,n} \tilde{x}_n \cos\left(w_n t\right)=\Re\big\{\sum_{n=0}^{N-1}\left\|\mathbf{h}_n\right\| s_{I,n} \tilde{x}_n e^{j w_n t}\big\}$.
Hence, with \eqref{opt_weight}, the multi-antenna multisine WIPT weight optimization is converted into an effective single antenna multi-carrier WIPT optimization with the effective channel gain on frequency $n$ given by $\left\|\mathbf{h}_n\right\|$ and the power allocated to the $n^{th}$ subband given by $s_{P,n}^2$ and $s_{I,n}^2$ for the multi-carrier unmodulated (multisine) and modulated waveform, respectively (subject to $\sum_{n=0}^{N-1}s_{P,n}^2+s_{I,n}^2=2P$). The optimum magnitude $s_{P,n}$ and $s_{I,n}$ in \eqref{opt_weight} can now be obtained by using the posynomial maximization methodology of Section \ref{PC_ID}. Namely, focusing on $n_o=4$ for simplicity, plugging \eqref{opt_weight} into \eqref{y_DC_2_2}, \eqref{y_DC_4_1b}, \eqref{E_y_I_2} and \eqref{E_y_I_4}, we get \eqref{y_DC_w_3} and \eqref{y_DC_w_4}.
\begin{table*}
\begin{align}
\mathcal{A}\left\{y_P(t)^2\right\}&=\frac{1}{2}\left[\sum_{n=0}^{N-1} \left\|\mathbf{h}_n\right\|^2 s_{P,n}^2\right],\mathcal{A}\left\{y_P(t)^4\right\}=\frac{3}{8}\left[\sum_{\mycom{n_0,n_1,n_2,n_3}{n_0+n_1=n_2+n_3}}\Bigg[\prod_{j=0}^3s_{P,n_j}\left\|\mathbf{h}_{n_j}\right\|\Bigg]\right],\label{y_DC_w_3}\\
\mathcal{E}\left\{\mathcal{A}\left\{y_I(t)^2\right\}\right\}&=\frac{1}{2}\left[\sum_{n=0}^{N-1} \left\|\mathbf{h}_n\right\|^2 s_{I,n}^2\right],\mathcal{E}\left\{\mathcal{A}\left\{y_I(t)^4\right\}\right\}=\frac{6}{8}\left[\sum_{n=0}^{N-1} \left\|\mathbf{h}_n\right\|^2 s_{I,n}^2\right]^2.\label{y_DC_w_4}
\end{align}\hrulefill
\end{table*}
The DC component $z_{DC}$ as defined in \eqref{z_DC_SWIPT} simply writes as $z_{DC}\left(\mathbf{s}_P,\mathbf{s}_I,\rho\right)$, expressing that it is now only a function of the $N$-dimensional vectors $\mathbf{s}_{P/I}=\left[\begin{array}{c} s_{P/I,0},\ldots,s_{P/I,N-1}\end{array}\right]$. 
Similarly, $I(\mathbf{s}_I,\rho)=\log_2\big(\prod_{n=0}^{N-1}\big(1+\frac{(1-\rho)}{\sigma_n^2} s_{I,n}^2\left\|\mathbf{h}_n\right\|^2\big)\big)$. 
\par Following the posynomial maximization methodology, we can write $z_{DC}(\mathbf{s}_P,\mathbf{s}_I,\rho)=\sum_{k=1}^K g_k(\mathbf{s}_P,\mathbf{s}_I,\rho)$ and $1+\frac{\bar{\rho}}{\sigma_n^2} C_n=\sum_{k=1}^{K_n}g_{nk}\left(\mathbf{s}_I,\bar{\rho}\right)$ with $C_n=s_{I,n}^2\left\|\mathbf{h}_n\right\|^2$, apply the AM-GM inequality and write the standard GP problem
\begin{align}\label{standard_GP_simple}
\min_{\mathbf{s}_P,\mathbf{s}_I,\rho,\bar{\rho},t_0} \hspace{0.3cm} &1/t_0\\
\textnormal{subject to} \hspace{0.3cm} &\frac{1}{2}\left[\left\|\mathbf{s}_P\right\|^2+\left\|\mathbf{s}_I\right\|^2\right]\leq P,\\ 
&t_0\prod_{k=1}^K\left(\frac{g_k(\mathbf{s}_P,\mathbf{s}_I,\rho)}{\gamma_k}\right)^{-\gamma_k}\leq1,\\
&2^{\bar{R}}\prod_{n=0}^{N-1}\prod_{k=1}^{K_n}\left(\frac{g_{nk}(\mathbf{s}_I,\bar{\rho})}{\gamma_{nk}}\right)^{-\gamma_{nk}}\leq 1,\label{standard_GP_constraint_3_simple}\\
&\rho+\bar{\rho}\leq 1.\label{standard_GP_SWIPT_4_simple}
\end{align}

Algorithm \ref{Algthm_OPT_simple} summarizes the design methodology with spatial and frequency domain decoupling. Such an approach would lead to the same performance as the joint space-frequency design of Algorithm \ref{Algthm_OPT_WIPT} but would significantly reduce the computational complexity since only $N$-dimensional vectors $\mathbf{s}_P$ and $\mathbf{s}_I$ are to be optimized numerically, compared to the $N\times M$ matrices $\mathbf{S}_P$ and $\mathbf{S}_I$ of Algorithm \ref{Algthm_OPT_simple}.
\begin{algorithm}
\caption{WIPT Waveform with Decoupling}
\label{Algthm_OPT_simple}
\begin{algorithmic}[1]
\State \textbf{Initialize}: $i\gets 0$, $\bar{R}$, $\mathbf{w}_{P/I,n}$ and in \eqref{opt_weight}, $\mathbf{s}_P$, $\mathbf{s}_I$, $\rho$, $\bar{\rho}=1-\rho$, $z_{DC}^{(0)}=0$
\label{Algthm_OPT_WIPT_step_initialize_simple}
\Repeat
    \State $i\gets i+1$, $\ddot{\mathbf{s}}_P\gets \mathbf{s}_P$, $\ddot{\mathbf{s}}_I\gets \mathbf{s}_I$, $\ddot{\rho}\gets \rho$, $\ddot{\bar{\rho}}\gets \bar{\rho}$
    \State $\gamma_k\gets g_k(\ddot{\mathbf{s}}_P,\ddot{\mathbf{s}}_I,\ddot{\rho})/z_{DC}(\ddot{\mathbf{s}}_P,\ddot{\mathbf{s}}_I,\ddot{\rho})$, $k=1,\ldots,K$  
    \label{Algthm_OPT_WIPT_step_gamma_simple}
    \State $\gamma_{nk}\gets g_{nk}(\ddot{\mathbf{s}}_I,\ddot{\bar{\rho}})/\big(1+\frac{\ddot{\bar{\rho}}}{\sigma_n^2}C_n(\ddot{\mathbf{s}}_I)\big)$, $n=0,\ldots,N-1$, $k=1,\ldots,K_{n}$  
    \label{Algthm_OPT_WIPT_step_gamma_2_simple}
    \State  $\mathbf{s}_P,\mathbf{s}_I,\rho,\bar{\rho} \gets \arg \min \eqref{standard_GP_simple}-\eqref{standard_GP_SWIPT_4_simple}$
    \label{Algthm_OPT_WIPT_step_OPT_simple}
    \State $z_{DC}^{(i)} \gets z_{DC}(\mathbf{s}_P,\mathbf{s}_I,\rho)$
\Until{$\left|z_{DC}^{(i)} - z_{DC}^{(i-1)} \right| < \epsilon$ \text{or} $i=i_{\max}$ }
\end{algorithmic}
\end{algorithm}

\vspace{-0.1cm}
\subsection{WIPT: Characterizing the Twofold Benefit of using Deterministic Power Waveforms}\label{NC_ID}

\par Superposing a power waveform to a communication waveform may boost the energy performance but may lead to the drawback that the power waveform interferes with the communication waveform at the information receiver. Uniquely, with the proposed architecture, this issue does not occur since the power waveform is deterministic, which leads to this twofold energy and rate benefit highlighted in Remark \ref{twofold_benefit}. Nevertheless, in order to get some insights into the contributions of the deterministic power waveform to the twofold benefit, we compare in Section \ref{simulations} the R-E region of Section \ref{PC_ID} to that of two baselines. The first baseline is the simplified architecture where the deterministic multisine waveform is absent and for which the R-E region can still be computed using Algorithm \ref{Algthm_OPT_WIPT} by forcing $\mathbf{S}_P=\mathbf{0}$. In this setup, the twofold benefit disappears. The second baseline is the hypothetical system where the power waveform is a deterministic multisine from an energy perspective but CSCG distributed from a rate perspective. The energy benefit is retained but the rate benefit disappears since the power waveform now creates an interference term $\sqrt{1-\rho}\mathbf{h}_n\mathbf{w}_{P,n}$ (with power level given by $(1-\rho)\left|\mathbf{h}_n\mathbf{w}_{P,n}\right|^2$) at the information receiver and therefore a rate loss. The achievable rate of that system is given by
\begin{multline}
I_{LB}(\mathbf{S}_P,\mathbf{S}_I,\mathbf{\Phi}_P,\mathbf{\Phi}_I,\rho)\\=\sum_{n=0}^{N-1} \log_2\left(1+\frac{(1-\rho)\left|\mathbf{h}_n\mathbf{w}_{I,n}\right|^2}{\sigma_n^2+(1-\rho)\left|\mathbf{h}_n\mathbf{w}_{P,n}\right|^2} \right)\label{R_NC}
\end{multline}
and the corresponding R-E region is a lower bound on that achieved in Section \ref{PC_ID}. Comparing those two R-E regions, we get a sense of the rate benefit of using a deterministic power waveform over a modulated one.

\par The lower bound on the R-E region is obtained by replacing $I$ by $I_{LB}$ in \eqref{R_zDC_region}. Because of the interference term in the SINR expression, decoupling the space frequency design by choosing the weights vectors as in \eqref{opt_weight} is not guaranteed to be optimal. To characterize the R-E region of the hypothetical system, we therefore resort to the more general approach where space and frequency domain weights are jointly designed, similarly to the one used in Section \ref{PC_ID}. 
\par Let us assume the same phases $\mathbf{\Phi}_P^{\star}$ and $\mathbf{\Phi}_I^{\star}$ as in \eqref{opt_phases}. Such a choice is optimal for $M=1$ (even though there is no claim of optimality for $M>1$). With such a choice of phases, $z_{DC}(\mathbf{S}_P,\mathbf{S}_I,\mathbf{\Phi}_P^{\star},\mathbf{\Phi}_I^{\star},\rho)$ also writes as \eqref{z_DC_SWIPT_final}, while the lower bound on the achievable rate is written as
\begin{equation}
I_{LB}(\mathbf{S}_I,\mathbf{\Phi}_I^{\star},\rho)=\log_2\left(\prod_{n=0}^{N-1}\left(1\!+\!\frac{(1\!-\!\rho)C_n}{\sigma_n^2\!+\!(1\!-\!\rho)D_n} \right)\right)\label{R_SWIPT_final_NC} 
\end{equation}
where $C_n=\sum_{m_0,m_1} \prod_{j=0}^1 s_{I,n,m_j}A_{n,m_j}$ and $D_n=\sum_{m_0,m_1} \prod_{j=0}^1 s_{P,n,m_j}A_{n,m_j}$.
\par The optimization problem can now be written as \eqref{SWIPT_opt_problem_eq}-\eqref{SWIPT_opt_problem_eq_4} with \eqref{SWIPT_opt_problem_eq_4} replaced by
\begin{equation}
2^{\bar{R}}\frac{\prod_{n=0}^{N-1}\left(1+\frac{\bar{\rho}}{\sigma_n^2}D_n\right)}{\prod_{n=0}^{N-1}\left(1+\frac{\bar{\rho}}{\sigma_n^2}\left(D_n+C_n\right)\right)}\leq 1.\label{SWIPT_opt_problem_eq_4_NC}
\end{equation}

\par Define the set of monomials $\left\{f_{nj}(\mathbf{S}_P,\mathbf{S}_I,\rho)\right\}$ of the posynomial $1\!+\!\frac{\bar{\rho}}{\sigma_n^2}\left(D_n\!+\!C_n\right)\!=\!\sum_{j=1}^{J_n}f_{nj}(\mathbf{S}_P,\mathbf{S}_I,\rho)$. For a given choice of $\left\{\gamma_{nj}\right\}$ with $\gamma_{nj}\geq 0$ and $\sum_{j=1}^{J_n} \gamma_{nj}\!=\!1$, we perform single condensations and write the standard GP as
\begin{align}\label{standard_GP_SWIPT_NC}
\min_{\mathbf{S}_P,\mathbf{S}_I,\rho,\bar{\rho},t_0} \hspace{0.3cm} &1/t_0\\
\textnormal{subject to} \hspace{0.3cm} &\frac{1}{2}\big[\left\|\mathbf{S}_I\right\|_F^2+\left\|\mathbf{S}_P\right\|_F^2\big]\leq P,\\ 
&t_0\prod_{k=1}^K\left(\frac{g_k(\mathbf{S}_P,\mathbf{S}_I,\mathbf{\Phi}_P^{\star},\mathbf{\Phi}_I^{\star},\rho)}{\gamma_k}\right)^{-\gamma_k}\leq1,\\
&2^{\bar{R}}\prod_{n=0}^{N-1}\left(1+\frac{\bar{\rho}}{\sigma_n^2}D_n(\mathbf{S}_P)\right)\nonumber\\
&\hspace{1cm}\prod_{j=1}^{J_n}\left(\frac{f_{nj}(\mathbf{S}_P,\mathbf{S}_I,\rho)}{\gamma_{nj}}\right)^{-\gamma_{nj}}\leq 1,\\
&\rho+\bar{\rho}\leq 1.\label{standard_GP_SWIPT_4_NC}
\end{align}
The whole optimization procedure is summarized in Algorithm \ref{Algthm_OPT_WIPT_NC}. Algorithm \ref{Algthm_OPT_WIPT_NC} boils down to Algorithm \ref{Algthm_OPT_WIPT} for $D_n=0$ $\forall n$.

\begin{algorithm}
\caption{Lower-Bound on R-E Region}
\label{Algthm_OPT_WIPT_NC}
\begin{algorithmic}[1]
\State \textbf{Initialize}: $i\gets 0$, $\bar{R}$, $\mathbf{\Phi}_P^{\star}$ and $\mathbf{\Phi}_I^{\star}$, $\mathbf{S}_P$, $\mathbf{S}_I$, $\rho$, $\bar{\rho}=1-\rho$, $z_{DC}^{(0)}=0$
\label{Algthm_OPT_WIPT_step_initialize_NC}
\Repeat
    \State $i\gets i+1$, $\ddot{\mathbf{S}}_P\gets \mathbf{S}_P$, $\ddot{\mathbf{S}}_I\gets \mathbf{S}_I$, $\ddot{\rho}\gets \rho$, $\ddot{\bar{\rho}}\gets \bar{\rho}$
    \State $\gamma_k\gets g_k(\ddot{\mathbf{S}}_P,\ddot{\mathbf{S}}_I,\mathbf{\Phi}_P^{\star},\mathbf{\Phi}_I^{\star},\ddot{\rho})/z_{DC}(\ddot{\mathbf{S}}_P,\ddot{\mathbf{S}}_I,\mathbf{\Phi}_P^{\star},\mathbf{\Phi}_I^{\star},\ddot{\rho})$, $k=1,\ldots,K$  
    \label{Algthm_OPT_WIPT_step_gamma_NC}
    \State $\gamma_{nj}\gets f_{nj}(\ddot{\mathbf{S}}_P,\ddot{\mathbf{S}}_I,\ddot{\rho})/\big(1+\frac{\ddot{\bar{\rho}}}{\sigma_n^2}(D_n(\ddot{\mathbf{S}}_P)+C_n(\ddot{\mathbf{S}}_I))\big)$, $n=0,\ldots,N-1$, $j=1,\ldots,J_{n}$  
    \label{Algthm_OPT_WIPT_step_gamma_2_NC}
    \State  $\mathbf{S}_P,\mathbf{S}_I,\rho,\bar{\rho} \gets \arg \min \eqref{standard_GP_SWIPT_NC}-\eqref{standard_GP_SWIPT_4_NC}$
    \label{Algthm_OPT_WIPT_step_OPT_NC}
    \State $z_{DC}^{(i)} \gets z_{DC}(\mathbf{S}_P,\mathbf{S}_I,\mathbf{\Phi}_P^{\star},\mathbf{\Phi}_I^{\star},\rho)$
\Until{$\left|z_{DC}^{(i)} - z_{DC}^{(i-1)} \right| < \epsilon$ \text{or} $i=i_{\max}$ }
\end{algorithmic}
\end{algorithm}

\vspace{-0.3cm}
\section{Scaling Laws}\label{scaling_laws}
\par In order to get some insights and assess the performance gain/loss of an unmodulated waveform over a modulated waveform for WPT, we quantify how $z_{DC}$ scales as a function of $N$ for $M=1$. For simplicity we truncate the Taylor expansion to the fourth order ($n_o=4$). We consider frequency-flat (FF) and frequency-selective (FS) channels. We assume that the complex channel gains $\alpha_l e^{j\xi_l}$ are modeled as independent CSCG random variables. $\alpha_l$ are therefore independent Rayleigh distributed such that $\alpha_l^2\sim \textnormal{EXPO}(\lambda_l)$ with $1/\lambda_l=\beta_l=\mathcal{E}\left\{\alpha_l^2\right\}$. The impulse responses have a constant average received power normalized to 1 such that $\sum_{l=0}^{L-1}\beta_l=1$. In the frequency flat channel, $\bar{\psi}_{n}=\bar{\psi}$ and $A_n=A$ $\forall n$. This is met when $(N-1)\Delta_f$ is much smaller than the channel coherence bandwidth. In the frequency selective channel, we assume that $L>>1$ and frequencies $f_n$ are far apart from each other such that the frequency domain CSCG random channel gains $h_{n,m}$ fade independently (phase and amplitude-wise) across frequencies and antennas. Taking the expectation over the distribution of the channel, we denote $\bar{z}_{DC}=\mathcal{E}\left\{z_{DC}\right\}$. Leveraging the derivation of the scaling laws for multisine waveforms in \cite{Clerckx:2016b}, we can compute the scaling laws of multi-carrier modulated waveforms designed using the ASS strategy (relying on CSIT) and the uniform power (UP) allocation strategy (not relying on CSIT). UP is characterized by $\mathbf{S}_i=\sqrt{2P}/\sqrt{N}\mathbf{1}_N$ and $\mathbf{\Phi}_i=\mathbf{0}_N$, $i\in\left\{P,I\right\}$ \cite{Clerckx:2016b}.
\begin{table*}
\caption{Scaling Laws of Multi-Carrier Unmodulated vs Modulated Waveforms.}
\centering
\begin{tabular}{|p{1.5cm}|p{1.5cm}|p{2cm}||p{4cm}|p{6cm}|}
\hline \textbf{Waveform} & \textbf{Strategy} & $N,M$	& \textbf{Frequency-Flat (FF)} &	\textbf{Frequency-Selective (FS)} \\
\hline
\hline \textbf{No CSIT} &	&  &  &  \\
\hline  Modulated  & $\bar{z}_{DC,SC}$	& $N\! =\! 1,M\!=\!1$ & $k_2 R_{ant} P+6 k_4 R_{ant}^2 P^2$	&  \\
\hline  Modulated  & $\bar{z}_{DC,UP}$	& $N\! >\! 1,M\!=\!1$ & $k_2 R_{ant} P+6 k_4 R_{ant}^2 P^2$	& $k_2R_{ant}P + 6k_4 R_{ant}^2 P^2$ \\
\hline  Unmodulated	& $\bar{z}_{DC,SC}$ & $N\! =\! 1,M\!=\!1$ & $k_2 R_{ant} P+3 k_4 R_{ant}^2 P^2$	&  \\
\hline  Unmodulated	& $\bar{z}_{DC,UP}$ & $N\! >>\! 1,M\!=\!1$ & $k_2 R_{ant} P+2 k_4 R_{ant}^2 P^2 N$	& $k_2R_{ant}P + 3k_4 R_{ant}^2 P^2$ \\
\hline \textbf{CSIT} &	&  &  &  \\
\hline  Modulated  & $\bar{z}_{DC,ASS}$	& $N\! >>\! 1,M\!=\!1$ & $k_2 R_{ant} P+6 k_4 R_{ant}^2 P^2$	& $k_2R_{ant}P \log N + 3 k_4 R_{ant}^2 P^2 \log^2 N$ \\
\hline  Unmodulated	& $\bar{z}_{DC,ASS}$ & $N\! >>\! 1,M\!=\!1$ & $k_2 R_{ant} P+3 k_4 R_{ant}^2 P^2$	& $k_2R_{ant}P \log N + \frac{3}{2}k_4 R_{ant}^2 P^2 \log^2 N$ \\
\hline  Unmodulated & $\bar{z}_{DC,UPMF}$ &  $N\! >>\! 1,M\! =\! 1$ & $k_2 R_{ant} P+2k_4 R_{ant}^2 P^2 N $	& $\geq k_2 R_{ant} P+\pi^2/16 k_4 R_{ant}^2 P^2 N$ \hspace{4cm} $\leq k_2 R_{ant} P +2 k_4 R_{ant}^2 P^2 N$ \\
\hline
\end{tabular}
\label{scaling_law_summary}
\vspace{-0.3cm}
\end{table*}

\par Table \ref{scaling_law_summary} summarizes the scaling laws for both modulated and unmodulated waveforms for $N=1$ (single-carrier SC) and for $N>1$ (multi-carrier) with ASS and UP strategies. It also compares with the UPMF strategy for multi-carrier unmodulated (multisine) waveforms (introduced in \cite{Clerckx:2016b}) that consists in uniformly allocating power to all sinewaves and matching the waveform phase to the channel phase (hence it requires CSIT). Such a UPMF strategy is suboptimal for multisine excitation with the nonlinear model-based design and its scaling law is therefore a lower bound on what can be achieved with the optimal multisine strategy of \eqref{WPT_max} \cite{Clerckx:2016b}. 
\par Let us first discuss \textit{multi-carrier transmission} ($N>>1$) \textit{with CSIT}. Recall that, in the presence of CSIT, the ASS strategy for multi-carrier modulated waveforms is optimal for the maximization of $z_{DC}$ with the linear and nonlinear model-based designs. The ASS for unmodulated (multisine) waveform is only optimal for the linear model-based design. As it can be seen from the scaling laws\footnote{Taking $N$ to infinity does not imply that the harvested energy reaches infinity as explained in Remark 4 of \cite{Clerckx:2016b}.}, unmodulated (multisine) waveform with UPMF strategy leads to a linear increase of $\bar{z}_{DC}$ with $N$ in FF and FS channels while the ASS strategy for both modulated and unmodulated only lead to at most a logarithmic increase with $N$ (achieved in FS channels). Hence, despite being suboptimal for the nonlinear model-based design, an unmodulated (multisine) waveform with UPMF provides a better scaling law than those achieved by modulated/unmodulated waveforms with ASS. In other words, even a suboptimal multi-carrier unmodulated waveform outperforms the optimal design of the multi-carrier modulated (with CSCG inputs) waveform.
\par Let us now look at \textit{multi-carrier transmission in the absence of CSIT}. A modulated waveform does not enable a linear increase of $\bar{z}_{DC}$ with $N$ in FF channels, contrary to the unmodulated waveform. This is due to the CSCG distributions of the information symbols that create random fluctuations of the amplitudes and phases across frequencies. This contrasts with the periodic behavior of the unmodulated multisine waveform that has the ability to excite the rectifier in a periodic manner by sending higher energy pulses every $1/\Delta_f$. 
\par Finally, for the \textit{single-carrier transmission in the absence of CSIT} case ($N=1$), an opposite behavior is observed with $\bar{z}_{DC}$ of the modulated waveform outperforming that of the unmodulated waveform thanks to the fourth order term being twice as large. This gain originates from the presence of the fourth order moment of $\tilde{x}_n$, namely $\mathcal{E}\big\{\left|\tilde{x}_n\right|^4\big\}=2$, in the fourth order term of \eqref{diode_model_2_OFDM}. Hence modulation through the CSCG distribution of the input symbols is actually beneficial to WPT in single-carrier transmissions. Actually, this also suggests that a modulation with an input distribution leading to a large fourth order moment is beneficial to WPT \footnote{Asymmetric Gaussian signaling (asymmetric power allocations to the real and imaginary dimensions) may be a better option than CSCG for WPT \cite{Varasteh:2017}.}.
\par The scaling laws also highlight that the effect of channel frequency selectivity on the performance depends on the type of waveform. Indeed, without CSIT, frequency selectivity is detrimental to unmodulated waveform performance (as already confirmed in \cite{Clerckx:2016b}) but has no impact on modulated waveform performance. On the other hand, with CSIT, frequency selectivity leads to a frequency diversity gain that is helpful to the performance of both types of waveforms\footnote{This behavior was already confirmed in \cite{Clerckx:2016b} for multisine waveforms.}. 

\begin{observation}\label{obs_main_2} The scaling laws highlight that, due to the diode nonlinearity of the fourth order term, there is a clear difference between using an unmodulated (multisine) waveform and a modulated waveform in WPT. For \textit{single-carrier transmissions} ($N=1$), the scaling law of the modulated waveform outperforms that of the unmodulated waveform due to the beneficial effect of the fourth order moment of the CSCG distribution to boost the fourth order order term. On the other hand, for \textit{multi-carrier transmissions} ($N>>1$), the scaling law of multisine significantly outperforms that of the modulated waveform. While $\bar{z}_{DC}$ scales linearly with $N$ with a multisine waveform thanks to all carriers being in-phase, it scales at most logarithmically with $N$ with the modulated waveform due to the independent CSCG randomness of the information symbols across subbands. This also shows that the diode nonlinearity can be beneficial to WPT performance in two different ways: first, by involving higher order moments of the input distribution, and second by enabling constructive contributions from various frequencies.
\end{observation}

Observation \ref{obs_main_2} further motivates the WIPT architecture of Section \ref{SWIPT_section} that is based on a superposition of multi-carrier unmodulated (deterministic multisine) waveform for efficient WPT and multi-carrier modulated waveform for efficient WIT. 

\vspace{-0.2cm}
\section{Performance Evaluations}\label{simulations}

\par We consider two types of performance evaluations, the first one is based on the simplified nonlinear model of Section \ref{section_EH} with $n_o=4$, while the second one relies on an actual and accurate design of the rectenna in PSpice. 

\vspace{-0.1cm}
\subsection{Nonlinear Model-Based Performance Evaluations}

\begin{figure}
   \centerline{\includegraphics[width = 8cm]{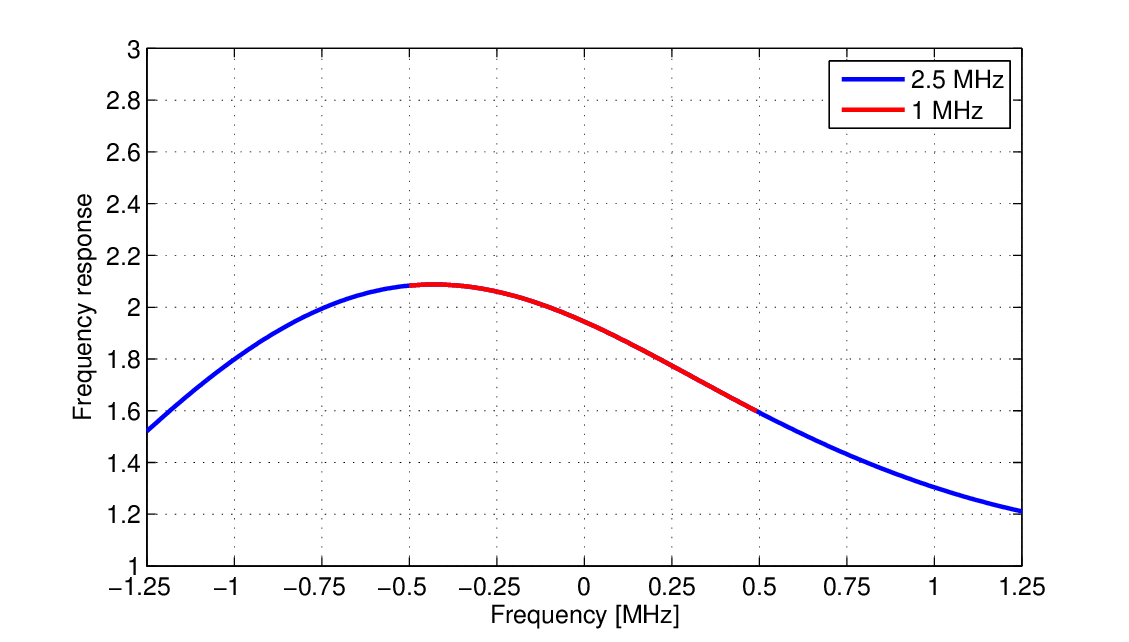}}
  \caption{Frequency response of the wireless channel.}
  \label{Freq_resp}
	\vspace{-0.4cm}
\end{figure}

\begin{figure*}
\begin{subfigmatrix}{1}
\subfigure[Superposed waveforms.]{\label{b}\includegraphics[width = 8cm]{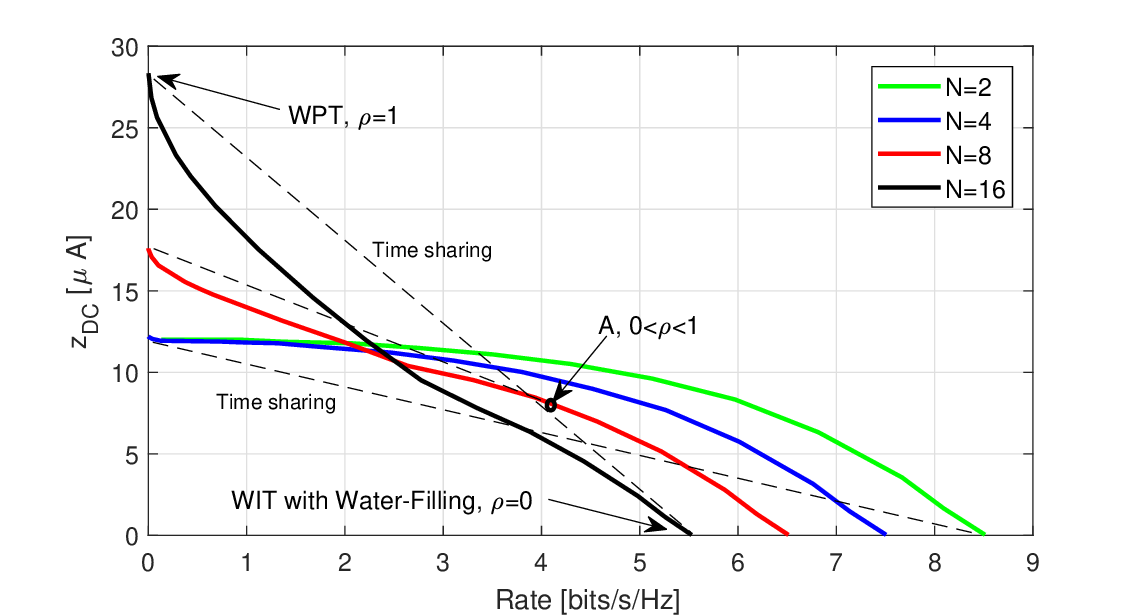}}
\subfigure[No power waveform.]{\label{d}\includegraphics[width = 8cm]{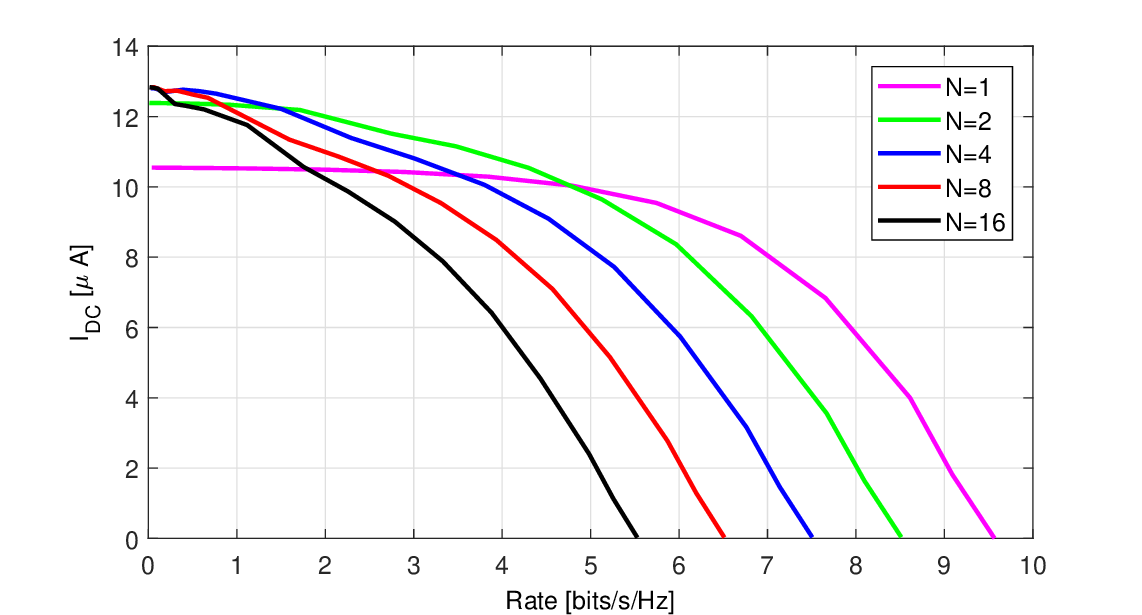}}
\end{subfigmatrix}
\caption{$C_{R-I_{DC}}$ as a function of $N$ for $M=1$, $B=1$MHz and SNR=20dB.}
\label{z_DC_results_SWIPT_N}
\end{figure*}

\par We now illustrate the performance of the optimized WIPT architecture. Parameters are taken as $k_2=0.0034$, $k_4=0.3829$ and $R_{ant}=50\Omega$. We assume a WiFi-like environment at a center frequency of 5.18GHz with a 36dBm EIRP, 2dBi receive antenna gain and 58dB path loss. This leads to an average received power of about -20dBm. We assume a large open space environment with a NLOS channel power delay profile with 18 taps obtained from model B \cite{Medbo:1998b}. Taps are modeled as i.i.d. CSCG random variables, each with an average power $\beta_l$. The multipath response is normalized such that $\sum_{l=1}^{18} \beta_l=1$. Equivalently, this system model can be viewed as a transmission over the aforementioned normalized multipath channel with an average transmit power fixed to -20dBm. Fig.\ \ref{Freq_resp} illustrates the frequency response of one realization of the multipath channel. 
\par The superposed multi-carrier modulated and unmodulated waveforms are designed assuming a frequency gap $\Delta_f\!=\!B/N$ with $B\!=\!1MHz$, and centered around 5.18GHz. 
\par Fig \ref{z_DC_results_SWIPT_N}(a),(b) illustrate the rate-energy (R-E) region obtained with superposed waveforms (with Algorithm \ref{Algthm_OPT_WIPT}) and without power (multisine) waveform (with Algorithm \ref{Algthm_OPT_WIPT} by allocating zero power to the unmodulated multisine waveform). $M=1$ and $N=1,2,4,8,16$ are considered and transmission is performed over the frequency response of Fig \ref{Freq_resp}. The noise power $\sigma_n^2$ is fixed at -40dBm in each subband, corresponding to a SNR (defined as $P/\sigma_n^2$) of 20dB. For the two strategies (superposed waveforms and no power waveform), the extreme right point on the x-axis refers to the rate achieved by the water-filling solution (with all transmit power allocated to the multi-carrier communication waveform) with $\rho=0$. Note that in all the figures, the rate has been normalized w.r.t.\ the bandwidth $N B_s$. Hence, the x-axis refers to a \textit{per-subband rate}. It decreases as $N$ increases because each subband receives a fraction of the total power. In Fig \ref{z_DC_results_SWIPT_N}(a), the points on the y-axis refer to the $z_{DC}$ achieved with the multisine waveform (WPT only) of \cite{Clerckx:2016b} with $\rho\!=\!1$ for $N$ sufficiently large (e.g.\ 8,16). 

\par A \textit{first} observation is that with superposed waveforms, increasing $N$ boosts the harvested energy. On the other hand, in the absence of multisine waveform, the harvested energy does not change much. This was predicted from the scaling laws in Table \ref{scaling_law_summary} that showed that the average $z_{DC}$ with a multi-carrier modulated waveform is independent of $N$ in a frequency-flat channel and only increases logarithmically with $N$ in a frequency-selective channel (as a consequence of a frequency diversity gain), and is lower than that achieved with the multisine waveform (which scaled linearly with $N$). We also note that the benefit of using superposed waveforms really kicks in for $N$ typically larger than 4. For $N$ smaller, there is no incentive to use a superposed waveform. This is also well aligned with the observations made from the scaling laws where a modulated waveform was outperforming the unmodulated waveform for small $N$ and inversely. Because of the frequency selectivity of the channel, a frequency diversity gain is exploited in Fig \ref{z_DC_results_SWIPT_N}. Its effect on $I_{DC}$ is for instance noticeable in Fig \ref{z_DC_results_SWIPT_N}(b) at very low rate. At such rate, the power is allocated to a single subcarrier/subband, namely the one corresponding to the strongest channel. For $N\!=\!1$, the subcarrier is located at the center frequency while for $N\!=\!2$, the two subcarriers (in baseband) are located at -0.25 and 0.25 MHz. At -0.25 MHz the channel gain is very close to the maximum experienced within the 1MHz band, which explains why the harvested energy for $N\!=\!2$ at low rate is only very slightly lower than that achieved for $N\!>\!2$. On the other hand, for $N\!=\!1$, the loss in harvested energy is more clearly visible due to the lower channel gain at the center frequency.

\par A \textit{second} observation is that without power/WPT waveform, the R-E region appears convex such that power-splitting (PS) outperforms time sharing (TS), irrespectively of $N$. On the other hand, with the superposed waveform, the R-E region with PS is convex for low $N$ ($N=2,4$) but concave-convex for larger $N$ ($N=8,16$). This has for consequences that PS-only is suboptimal. Looking at Fig.\ \ref{z_DC_results_SWIPT_N}(b) for $N=8$, the convex hull is obtained by TS between WPT (multisine waveform only and $\rho=1$) and WIPT with $0<\rho<1$ (illustrated by point A). Going for $N=16$, the convex hull is obtained by TS between the two extreme points: WPT (multisine waveform only and $\rho=1$) and WIT with water-filling (with $\rho=0$). This shows that the convex hull can in general be obtained by a combination of TS and PS. This behavior is observed because of the concacity-convexity of the R-E region, and fundamentally originates from the rectifier nonlinearity.
\par Note that those observations are in sharp contrast with the existing WIPT literature (relying on the linear model). It is indeed shown in \cite{Zhou:2013,Zhou:2014} that PS always outperforms TS.

\begin{figure*}
\begin{subfigmatrix}{2}
\subfigure[SNR=10dB.]{\label{a}\includegraphics[width = 8cm]{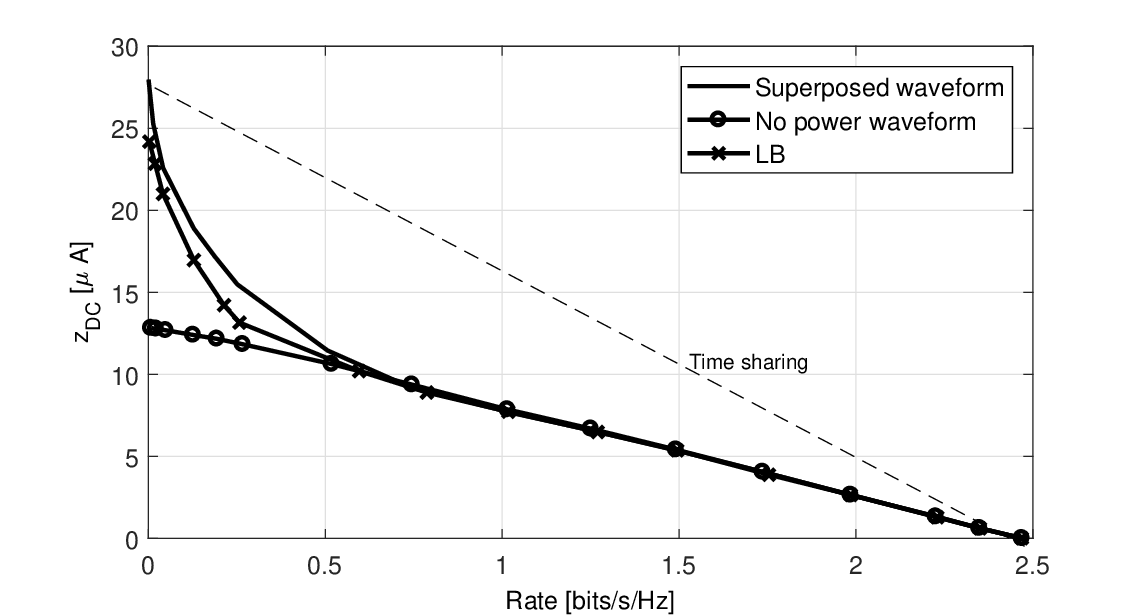}}
\subfigure[SNR=20dB.]{\label{b}\includegraphics[width = 8cm]{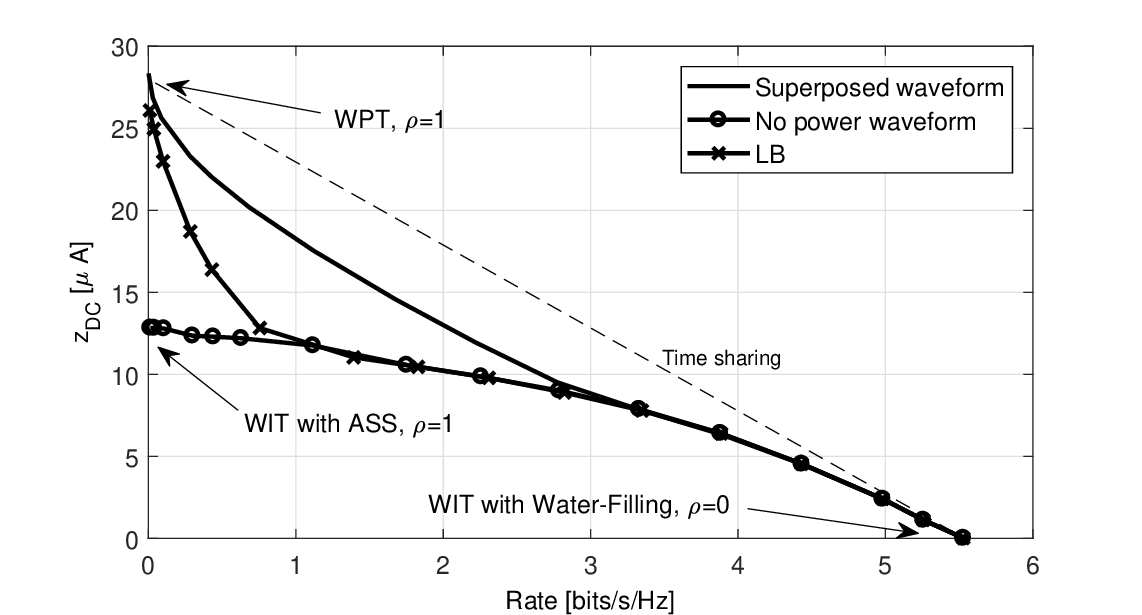}}
\subfigure[SNR=30dB.]{\label{b}\includegraphics[width = 8cm]{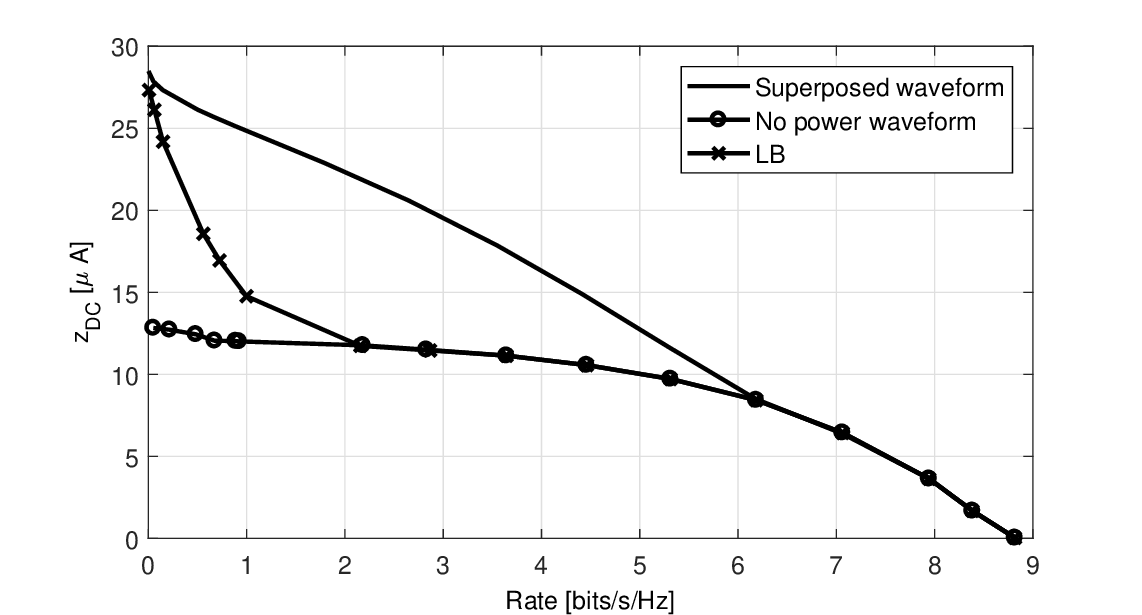}}
\subfigure[SNR=40dB.]{\label{b}\includegraphics[width = 8cm]{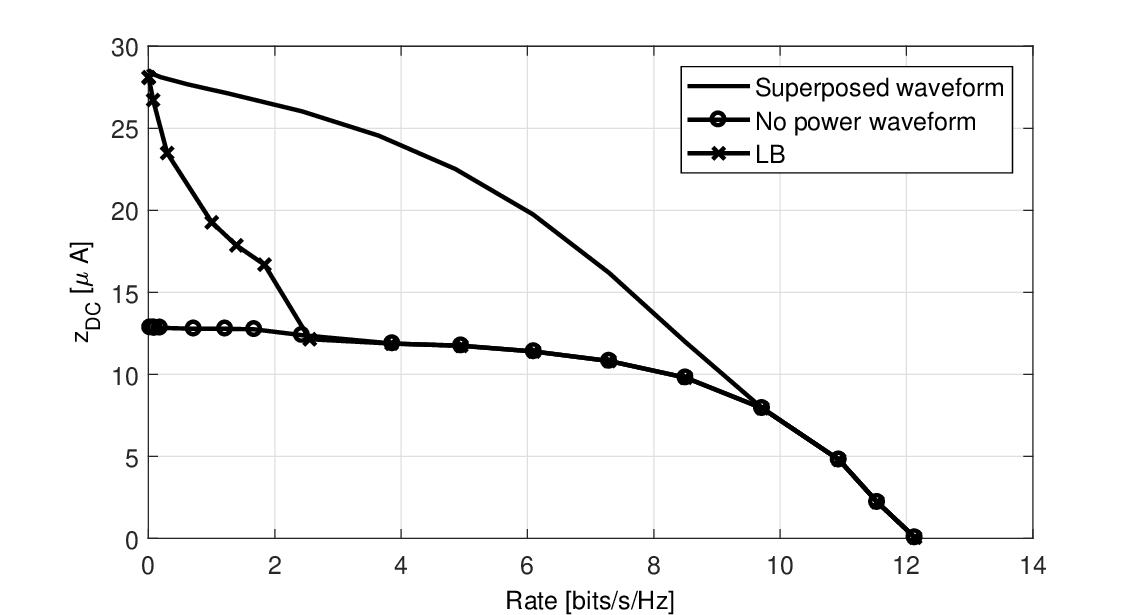}}
\end{subfigmatrix}
\caption{$C_{R-I_{DC}}$ for $N=16$ and $M=1$ over the multipath channel of Fig.\ \ref{Freq_resp} with $B=1$MHz and SNR=10,20,30,40 dB.}
\label{z_DC_results_SWIPT_SNR}
\vspace{-0.3cm}
\end{figure*}

\par Fig \ref{z_DC_results_SWIPT_SNR}(a),(b),(c),(d) further studies the performance of the WIPT architecture with superposed waveform for $N=16$ and four different SNRs (10,20,30,40 dB) under the same channel response as in Fig \ref{Freq_resp}. Comparisons are made with two baselines, namely the simplified system with ``no power waveform'' and the hypothetical system denoted by ``LB'' as described in Section \ref{NC_ID}. Comparing the R-E with superposed waveform to the ``no power waveform'' case, we gain insight into the joint twofold rate and energy benefit of using a deterministic power waveform. Comparing it to the the ``LB'' R-E region, we get insight into the rate benefit of using a power waveform that does not incur a rate at the information receiver.
\par A \textit{first} observation is that the R-E region with superposed waveform and its lower bound (LB) is augmented thanks to the presence of the deterministic multisine power waveform. This can be seen by noting that the corresponding R-E regions are larger than that obtained with no multisine waveform (No power waveform). This key observation fundamentally comes from the nonlinear behavior of the wireless power channel. The usefulness of the multisine originates from the fact that the rectifier is a nonlinear device and that the multisine is deterministic. On the contrary, the modulated waveform with CSCG inputs exhibits independent randomness of the information symbols across sub-carriers that leads to some uncontrollable fluctuation of the waveform at the input of the rectifier and therefore some loss in terms of harvested energy. Interestingly, this observation is in sharp contrast with the WIPT literature relying on the conventional linear model. It was indeed shown in \cite{Xu:2014} that, in the event where the power waveform is not eliminated (and therefore treated as noise) by the communication receiver, the power waveform is useless and does not help enlarging the R-E region. As we can see from Fig.\ \ref{z_DC_results_SWIPT_SNR} and Observation \ref{obs_main_2}, even in the worst-case ``LB'' where the power waveform is assumed to create interference to the information receiver, the multisine power waveform is always useful when the rectifier nonlinearity is properly taken into account in the design of WIPT. 
\par A \textit{second} observation is that the gain over ``LB'' increases as the SNR increases. At low SNR, the interference from the power waveform in ``LB'' is drawn within the noise and therefore does not impact much the R-E region. Both regions are therefore quite similar at very low SNR. On the other hand, as the SNR increases, in order for ``LB'' not to become interference limited, the hypothetical system of Section \ref{NC_ID} needs to allocate a very small power to the multisine waveform over a wide range of rates. For a given $z_{DC}$, this leads to a rate loss. This shows that the rate benefit of using a deterministic waveform increases as the SNR increases.
\par A \textit{third} observation is that at lower SNRs (10 and 20dB), the R-E region achieved superposed waveforms with PS is actually outperformed by a TS between WPT (multisine waveform only and $\rho=1$) and WIT (communication waveform only and $\rho=0$). At higher SNR (30 and 40dB SNR), PS performs better than TS between those two extreme points. This behavior is observed because of the concacity-convexity of the R-E region. On the other hand, in the absence of the WPT waveform (e.g.\ WIT-only transmission), the R-E region appears convex because of the inefficiency of the modulated waveform to boost $z_{DC}$. At low SNR (as on Fig \ref{z_DC_results_SWIPT_SNR}(a)), the water-filling (WF) strategy for rate maximization coincides with the ASS strategy for energy maximization. Hence the ASS strategy is used across the boundary and only changing $\rho$ enables to draw the region boundary. As the SNR increases and WF allocates power over an increasing number of subbands, the region boundary with ``No power waveform'' is getting rounded because of the log behavior of the rate expression.

\par In summary, we draw the following conclusions from Fig \ref{z_DC_results_SWIPT_N} and \ref{z_DC_results_SWIPT_SNR}: 1) A multi-carrier power (multisine) waveform (superposed to a multi-carrier communication waveform with CSCG inputs) is useful to enlarge the R-E region of WIPT, for $N$ sufficiently large ($N>4$). 2) Without multisine power waveform, PS is preferred over TS. 3) PS is favoured for low $N$, a combination of PS and TS for medium $N$ and TS for large $N$. 4) For sufficiently large $N$, TS is favoured at low SNR and PS at high SNR. 
It is important to note that 1), 3) and 4) are consequences of the non-linearity of the rectifier.

\par It is also noteworthy to recall that if we had used the linear model for the design and evaluation of the R-E region, following Observation \ref{obs_main}, there would not be any benefit of using a multisine waveform on top of the modulated waveform since both are equally suitable from an energy harvesting perspective. Moreover, according to Section \ref{OFDM_design}, the optimum design of a multi-carrier modulated waveform (with CSCG inputs) for WPT is the same for both the linear and non-linear model, namely based on the ASS strategy. These facts imply that the R-E region achieved with a design of WIPT based on the linear model is the same as the one achieved by ``No power waveform'' in Fig \ref{z_DC_results_SWIPT_N} and \ref{z_DC_results_SWIPT_SNR}\footnote{Validated by simulations but are not reported here for brevity.}. Moreover, since the ``No power waveform'' strategy is always significantly outperformed by the superposed waveform in Fig \ref{z_DC_results_SWIPT_N} and \ref{z_DC_results_SWIPT_SNR}, this also implies that the design of WIPT based on the nonlinear rectifier model is more spectrally efficient than a design based on the linear model. The importance of accounting for the nonlinearity of the rectifier in the design and evaluations of WPT was highlighted in \cite{Clerckx:2016b}. Evaluations in this paper shows that this is also true for WIPT.  

\par
\begin{figure}
\centerline{\includegraphics[width = 8cm]{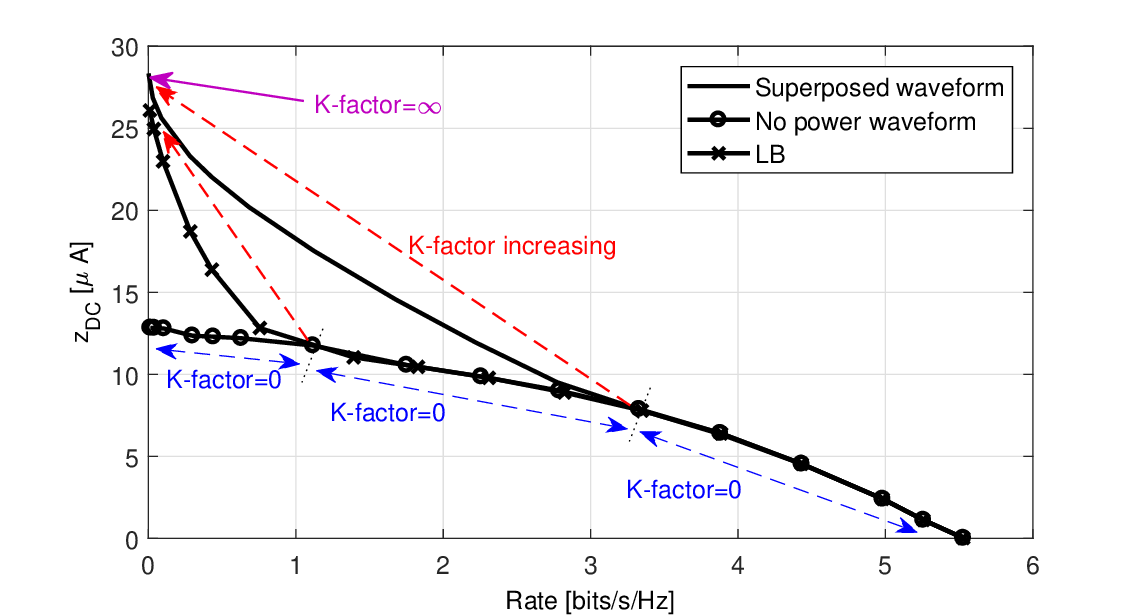}}
  \caption{$C_{R-I_{DC}}$ for $N=16$ and $M=1$ over the channel of Fig.\ \ref{Freq_resp} with $B=1$MHz and SNR=20dB. Evolution of the input distribution and K-factor.}
  \label{K_factor}
	\vspace{-0.3cm}
\end{figure}

 As a final but very interesting observation, we note that the input distribution in every subband with ``No power waveform'' or with a linear model-based design is CSCG. The mean is always zero and the variance is frequency-dependent. This is the classical capacity achieving input distribution over a Gaussian channel with an average transmit power constraint. Following Remark \ref{remark_distribution}, we note however that with the superposed waveform, the input distribution is Gaussian but can have a non-zero mean. It is zero-mean as long as only the multi-carrier modulated communication waveform (with CSCG inputs) is used, but as we reduce the threshold $\bar{R}$ and aim at a higher harvested energy, the waveform design algorithm starts allocating power over the multisine waveform. This has the effect of changing the input distribution from zero mean to non-zero mean. Concurrently the variance of the distribution decreases since the transmit power allocated to the modulated waveform is decreased, which leads to increasing K-factors (defined in Remark \ref{remark_distribution}). This also suggests an alternative way of interpreting the transmit waveform \eqref{SWIPT_WF}. Rather than viewing it as the superposition of a deterministic multisine and a modulated waveform with CSCG inputs, we can view it as a multi-carrier modulated waveform with a non-zero mean Gaussian input distribution. Fig \ref{z_DC_results_SWIPT_N} and \ref{z_DC_results_SWIPT_SNR} effectively show that designing a multi-carrier modulated waveform for WIPT with non-zero mean Gaussian inputs leads to an enlarged R-E region compared to the CSCG inputs. An interpretation of Fig \ref{z_DC_results_SWIPT_SNR}(b) in terms of K-factor is provided in Fig \ref{K_factor}. 

\vspace{-0.3cm}
\subsection{Validation of the Model and the Scaling Laws}\label{eval_circuit}
\par
\begin{figure}
\centerline{\includegraphics[width=0.8\columnwidth]{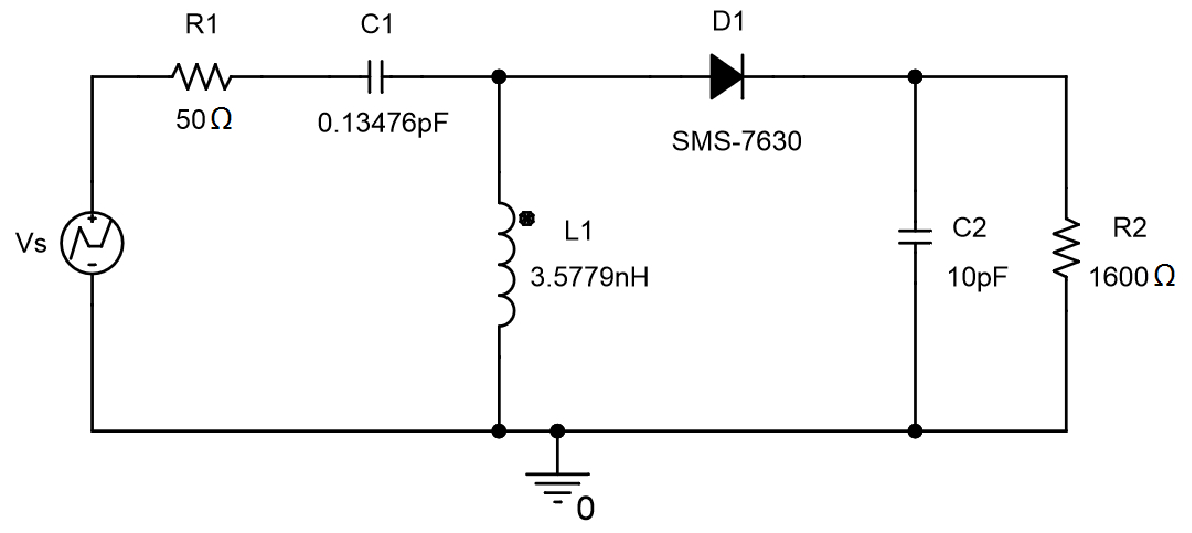}}
  \caption{Rectenna with a single diode and a L-matching network.}
  \label{circuit}
	\vspace{-0.4cm}
\end{figure}

\begin{figure}
\centerline{\includegraphics[width=0.9\columnwidth]{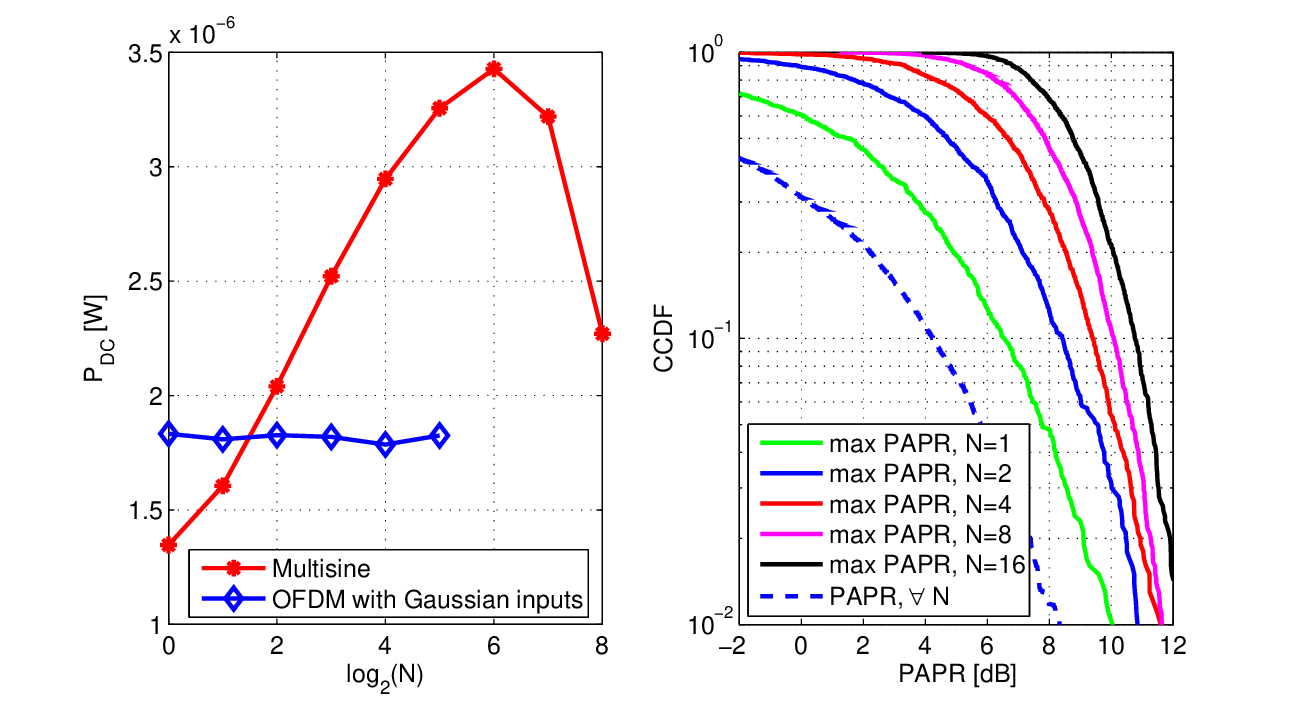}}
  \caption{$P_{dc}$ vs $N$ (left) and CCDF of PAPR with OFDM vs $N$ (right).}
  \label{Pdc_N}
\vspace{-0.3cm}
\end{figure}

 The rectifier model and the scaling laws for multisine were validated through circuit simulations using PSpice in \cite{Clerckx:2016b}. We here conduct a similar evaluation to validate the rectifier model and the scaling laws for modulated waveforms with CSCG inputs. To that end, PSpice simulations have been conducted using the realistic rectenna of Fig \ref{circuit} (same circuit as the one used in Fig 10 of \cite{Clerckx:2016b}) designed for an average input power of -20dBm (10$\mu$W). Details on the circuit design can be obtained in \cite{Clerckx:2016b}. 
Applying the UP strategy to both a multisine waveform and an OFDM waveform with CSCG inputs and assuming no wireless channel, i.e.\ $A\!=\!1$ and $\bar{\psi}\!=\!0$, Fig \ref{Pdc_N}(left) displays the harvested DC power $P_{dc}$ measured at the rectifier output load versus $N$, for $M=1$, $\Delta_f\!=\!B/N$ with $B\!=\!10$MHz and an OFDM symbol duration $T=1/\Delta_f$. 

\par The DC power of the OFDM waveform has a flat behavior as a function of $N$ while that of the multisine increases rapidly with $N$ (up to $N\!=\!64$)\footnote{For $N\!>\!64$, a decrease is observed because the rectenna has been optimized for $N\!=\!4$. Further explanations are provided in Fig 11 of \cite{Clerckx:2016b}.}. OFDM outperforms multisine for small $N$ and is outperformed for larger $N$. This validates the scaling laws of No CSIT in FF of Table \ref{scaling_law_summary} and Observation \ref{obs_main}. It also highlights the inaccuracy of the linear model that would have predicted that multisine and OFDM are equally suitable for WPT. The loss of OFDM compared to multisine for most $N\!>\!2$ comes from the random and independent fluctuations of the input symbols (magnitude and phase) across frequencies that lead to a random fluctuation of the input waveform, in contrast with the periodic behavior of the multisine waveform (which is more suitable to turn on and off the rectifier periodically). The gain for $N\! \leq\! 2$ comes from the fourth order moment of the CSCG distribution that boosts the fourth order term in $\bar{z}_{DC}$ by a factor 2 compared to the unmodulated case.
\par It is also worth contrasting with RF experiment results. In \cite{Sakaki:2014}, it was shown that (single-carrier) modulations such QPSK and 16QAM lead to amplitude and phases variations that are detrimental to the RF-to-DC conversion efficiency compared to a CW (i.e.\ N=1) in the input power range 0-10 dBm. This may appear contradicting our results since modulation hurts performance for $N=1$. Recall that our results are obtained with CSCG inputs, not with finite constellations.

\par The flat scaling law of the OFDM waveform may appear quite surprising. Recall that both OFDM and multisine waveforms exhibit a larger PAPR as $N$ increases, as illustrated in Fig \ref{Pdc_N}(right) for OFDM. We distinguish the CCDFs of ``max PAPR'' for $N=1,2,4,8,16$ and CCDF of ``PAPR $\forall N$''. The former ones are obtained by drawing the CCDF of $\max_{0\leq t \leq T} \left|x(t)\right|^2/\mathcal{E}\big\{\left|x(t)\right|^2\big\}$ (i.e.\ defined based on the maximum instantaneous power of the signal over a symbol duration). It is sensitive to $N$ because $\max_{0\leq t \leq T} \left|x(t)\right|^2$ is likely to increase as $N$ increases. On the contrary, the latter one displays the CCDF of $\left|x(t)\right|^2/\mathcal{E}\big\{\left|x(t)\right|^2\big\}$ (i.e.\ defined based on the instantaneous power of the signal). It is insensitive to $N$ because the modulated symbols are CSCG distributed and so is also the time domain signal irrespectively of $N$. For comparison, the PAPR of a multisine is given by $\max_{0\leq t \leq T} \left|x(t)\right|^2/\mathcal{E}\big\{\left|x(t)\right|^2\big\}=10\log(2N)$[dB], e.g. 6dB for $N\!=\!2$ and 9dB for $N\!=\!4$. Large PAPR has been advertised in the RF literature as a way to enhance the RF-to-DC conversion efficiency \cite{Trotter:2009,Boaventura:2011,Collado:2014,Valenta:2015}. With OFDM, in contrast with the multisine, $P_{dc}$ is insensitive to $N$ despite the increase of the (max) PAPR with $N$. Such a behavior of the OFDM waveform therefore cannot be explained by looking at PAPR. This shows that just PAPR is not an accurate enough metric to judge the suitability of a waveform for WPT. It should also be stressed that in \cite{Clerckx:2016b} (discussions along Fig 7 and 8 in that paper), the authors have already stressed and shown that maximizing PAPR is not a right approach to design efficient WPT multisine signals in frequency selective channels.

\begin{figure}
\centerline{\includegraphics[width=0.9\columnwidth]{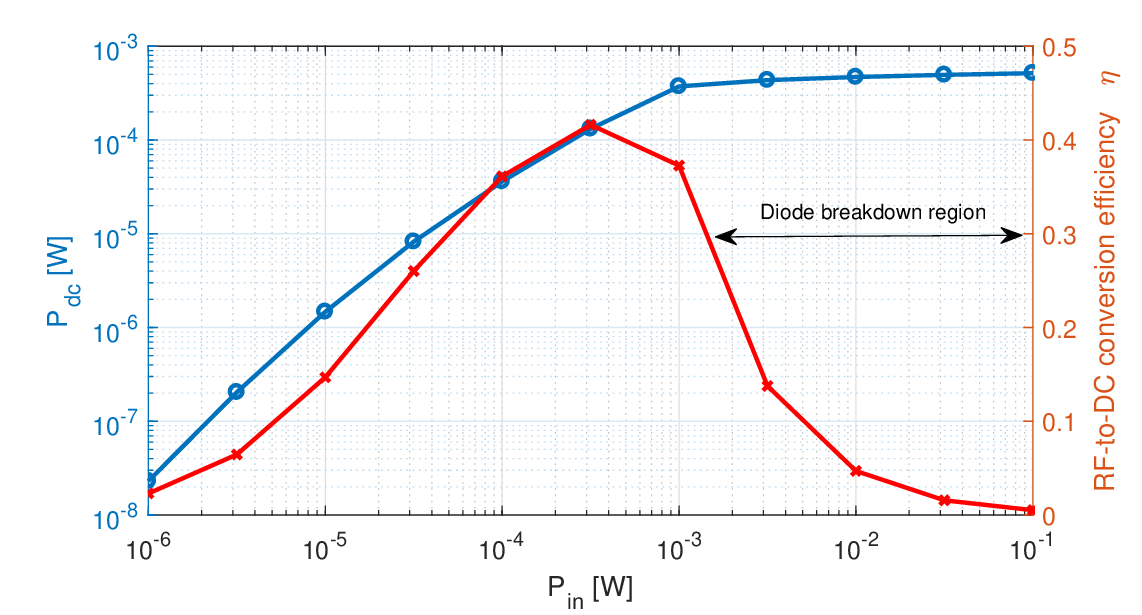}}
  \caption{$P_{dc}$ vs $P_{in}$ and RF-to-DC conversion efficiency $\eta$.}
  \label{Pdc_Pin}
\vspace{-0.3cm}
\end{figure}
\par In Fig \ref{Pdc_Pin}, we illustrate the discussion in Remark \ref{EH_remark}. We plot the DC power $P_{dc}$ harvested at the load of the circuit in Fig \ref{circuit} as a function of the input power $P_{in}$ to the rectifier when a CW (i.e.\ a single sinewave) signal is used for excitation. We also display the RF-to-DC conversion efficiency defined as $\eta=P_{dc}/P_{in}$. The diode SMS-7630 becomes reverse biased at 2Volts, corresponding to an input power of about 1mW. We note that beyond 1mW input power, the output DC power saturates and $\eta$ suddenly significantly drops, i.e.\ the rectifier enters the diode breakdown region. This circuit was designed for 10$\mu$W input power but as we can see it can operate typically between 1$\mu$W and 1mW. Beyond 1mW, a rectifier with multiple diodes will perform better and avoid the saturation problem \cite{OptBehaviour,Costanzo:2016,Sun:2013,Li:2014}.

\vspace{-0.2cm}
\section{Conclusions and Future Works}\label{conclusions}
The paper derived a methodology to design waveforms and characterize the rate-energy region of a point-to-point MISO WIPT. Contrary to the existing WIPT literature, the nonlinearity of the rectifier is modeled and taken into account in the WIPT waveform and transceiver optimization. Motivated by the fact that a multisine waveform outperforms a multi-carrier modulated waveform from a WPT perspective, a WIPT waveform is introduced as the superposition of a WPT waveform (multisine) and a WIT (multi-carrier modulated) waveform. The waveforms are adaptive to the CSI and result from a non-convex posynomial maximization problem.  
\par Results highlight that accounting for the rectifier nonlinearity radically changes the design of WIPT. It favours a different waveform, modulation, input distribution and transceiver architecture as well as a different use of the RF spectrum. Exploiting the rectifier nonlinearity in the WIPT design also makes a more efficient use of the resources by enabling enlarged rate-energy regions compared to those obtained by ignoring the nonlinearity in the system design.
\par A lot of interesting questions arise from this work. To name a few, a fundamental question is, how to make the best use of the RF spectrum for WIPT? Results here highlight that the conventional capacity-achieving CSCG input distribution is suboptimal for WIPT. What is the optimal input distribution and how to build corresponding WIPT architecture? Results also highlight that the design of WIPT differs from conventional communication due to the inherent nonlinearity of the wireless power channel. Interesting works consist in re-thinking WIPT architecture in light of nonlinearity for broadcast, multiple access, interference and relay channels. The problem of nonlinearity, waveform design and rate-energy tradeoff also occurs in other types of communication systems, such as in backscatter communications \cite{Clerckx:2017b}.

\ifCLASSOPTIONcaptionsoff
  \newpage
\fi

\vspace{-0.3cm}
\section{Acknowledgments}
We thank E. Bayguzina for providing Fig \ref{Pdc_N}(left) and Fig \ref{Pdc_Pin}.

\end{document}